\documentclass[prd,aps,twocolumn,superscriptaddress,preprintnumbers,nofootinbib,showpacs]{revtex4}
\usepackage{graphicx}
\usepackage{exscale}
\usepackage[intlimits]{amsmath}
\usepackage{amsfonts}
\usepackage{amssymb,amscd}
\usepackage{epsfig}
\usepackage{pstricks}

\newcounter{commentdepth}
\newcommand{\be}{\begin{equation}}
\newcommand{\ee}{\end{equation}}

\newcommand{\D}{\ensuremath\mathcal{D}}

\newcommand{\beqa}{\begin{eqnarray}}
\newcommand{\eeqa}{\end{eqnarray}}
\newcommand{\beq}{\begin{equation}}
\newcommand{\eeq}{\end{equation}}

\newcommand{\p}{\partial}

\newcommand{\reftitle}[1]{}

\begin{document}

\title{Goldstone bosons and fermions in QCD}


\author{Daniel~Zwanziger}
\affiliation{Physics Department, New York University, New York, NY 10003, USA}

\begin{abstract}
\noindent   	

We consider the version of QCD in Euclidean Landau gauge in which the restriction to the Gribov region is implemented by a local, renormalizable action.  This action depends on the Gribov parameter $\gamma$, with dimensions of (mass)$^4$, whose value is fixed in terms of $\Lambda_{QCD}$, by the gap equation, known as the horizon condition, ${\p \Gamma \over \p \gamma} = 0$, where $\Gamma$ is the quantum effective action.  The restriction to the Gribov region suppresses gluons in the infrared, which nicely explains why gluons are not in the physical spectrum, but this only makes more mysterious the origin of the long-range force between quarks.  In the present article  we exhibit the symmetries of $\Gamma$, and show that the solution to the gap equation, which defines the classical vacuum, spontaneously breaks some of the symmetries of $\Gamma$.  This implies the existence of massless Goldstone bosons and fermions that do not appear in the physical spectrum.  Some of the Goldstone bosons may be exchanged between quarks, and are candidates for a long-range confining force.  As an exact result we also find that in the infrared limit the gluon propagator vanishes like~$k^2$.

\end{abstract}


\maketitle

\section{Introduction}

	We are reasonably confident that the interactions of quarks and gluons are correctly described by the non-Abelian gauge theory known as QCD.  This confidence is based on the success of perturbative calculations at high energy and numerical calculations in lattice QCD.  However we lack a satisfactory continuum description the phases of QCD, comparable to the Higgs model of electro-weak interactions, and  it remains a challenge to understand the confinement of quarks and gluons at the non-perturbative level.  There are several suggestive scenarios, which involve the dual Meissner effect with condensation of magnetic monopoles, the maximal Abelian gauge, the maximal center gauge, and the Coulomb gauge.  There is also a scenario in Euclidean Landau gauge that originated with Gribov \cite{Gribov:1977wm} that is based on the insight that there exist Gribov copies --- that is to say gauge-equivalent configurations that nevertheless satisfy the  (Landau) gauge condition --- and moreover that the dynamics is strongly affected if one cuts off the Euclidean functional integral to avoid over-counting these copies.\footnote{For a connection between scenarios in center and abelian gauge and the Gribov scenario, see \cite{GOZ:2005}.}

	This cut-off is non-local in $A$-space.  However just as the non-local Faddeev-Popov determinant may be represented in a local action by means of Faddeev-Popov ghosts, so the non-local cut-off in $A$-space may be represented in a local action by means of additional fermionic and bosonic ghosts \cite{Zwanziger:1989mf, Zwanziger:1993}.  The renormalizibility of this local action has been established \cite{Zwanziger:1993, Schaden:1994}, and its consistency has been studied \cite{Sorella:2005, Gracey:0510, Gracey:0605, Dudal:0806, Dudal:0808, Sorella:0904, Dudal:0908, Gracey:0909, Vandersickel0910, Dudal:0911, Baulieu:2009, Dudal:1001}.
	
	Ideally one would like to cut off at the boundary of the fundamental modular region, $\Lambda$, which may be taken to be the set of absolute minima on each gauge orbit of the minimizing functional in $D$ Euclidean dimensions \cite{Nakajima:1978, Semenov:1982, Zwanziger:1982}
\beq
F_A(g) \equiv \int d^Dx \ |{^g}A|^2.
\eeq 
Here ${^g}A_\mu = g^{-1} A_\mu g + g^{-1} \p_\mu g$ is the transform of the non-abelian gauge potential $A_\mu(x)$ by the local gauge transformation $g(x)$.  However we lack an explicit description of $\Lambda$, and we instead integrate over the Gribov region, $\Omega$, which is the set of relative minima on each gauge orbit of the minimizing functional, a region that is convex and bounded in every direction \cite{Zwanziger:1982}.  The Gribov region is larger than the fundamental modular region, $\Lambda \subset \Omega$ and $\Lambda \neq \Omega$ \cite{Semenov:1982}.  However, since the integral over~$\Omega$ can be represented as a functional integral with  a local renormalizable action, it provides an interesting quantum field theory of non-abelian gauge fields which is worthy of study in its own right.  Moreover the restriction to $\Omega$ has dynamical consequences which substantiate the confinement scenario originally proposed by Gribov, so this provides a valuable pathway to confinement.  In the present work we shall derive some exact dynamical consequences starting from the local, renormalizable action by the operation of the Goldstone mechanism.

	The organization of this article is as follows.  In sect.~II we introduce the local action $S$ and the horizon condition, and we exhibit a BRST operator that is explicitly but softly broken.  The one-loop vacuum free energy is calculated in Appendix~B.  In sect.~III local sources and the quantum effective action $\Gamma$ are introduced, and we show that~$\Gamma$ enjoys the rich symmetries of the BRST-invariant part of~$S$, provided that the sources are suitably transformed.  In sect.~IV we show that the horizon condition spontaneously breaks the symmetries of $\Gamma$.  The Slavnov-Taylor identity is derived in sect.~V.  Sect.~VI is based on the results of Appendix A where the Ward identities corresponding to the equations of motion of the auxiliary ghosts are solved, as is the Ward identity corresponding to the integrated equation of motion of  the Fadeev-Popov ghost $c$.  This allows us to replace the quantum effective action by the reduced quantum effective action $\Gamma^*$, which depends only on a reduced set of variables.  In sect.~VII the Slavnov-Taylor identity satisfied by $\Gamma^*$ is derived.  In sect.~VIII the global Ward identities satisfied by $\Gamma^*$ are derived, which include super-symmetry transformations.  In sect.~IX the basic tensor invariants under these symmetries are constructed.  New on-shell symmetries are derived in sect.~X, and the surviving tensor invariants are constructed.  In sect.~XI, $\Gamma^*$ is expanded about the symmetry-breaking vacuum determined by the horizon condition, and the ``flat" directions are found.  In sect.~XII the leading derivative terms in the ``flat" directions are found.  In sect.~XIII the effective action in the infrared limit is exhibited.  In sect.~XIV the infrared limit of the propagators of the Goldstone particles is found.  In sect.~XV the infrared limit of the propagators of the gluon and the non-Goldstone ghosts is found.  In sect.~XVI the effective quark-quark interaction due to the exchange of one Goldstone particle is found.  Sect.~XVII presents an alternative interpretation of the results in terms of spontaneous breaking of the symmetry of the local Langrangian density.  Sect.~XVIII contains our concluding remarks.

\section{Local action and horizon condition}

	A local renormalizable action is defined by 
\beq
\label{fullaction}
S = S_0 + S_\gamma =  \int d^Dx \ ({\cal L}_0 + {\cal L}_{\gamma})
\eeq
where	
\beq
{\cal L}_0 = {\cal L}_{FP} + {\cal L}_{aux},
\eeq
and
\beq
{\cal L}_{FP} = (1/4) \ F_{\mu\nu}^2 + i \p_\mu b A_\mu - \p_\mu \bar c D_\mu c 
\eeq	
is the Faddeev-Popov Lagrangian density.  The Yang-Mills field tensor is written
\beq
F_{\mu \nu} = \p_\mu A_\nu - \p_\nu A_\mu + g A_\mu \times A_\nu,
\eeq
where $(A_\mu \times A_\nu)^a \equiv f^{abc}A_\mu^b A_\nu^c$.  The Lagrange-multiplier field $b$ imposes the Landau gauge condition $\p \cdot A = 0$.  Without loss of generality we ignore the quark action which plays no role in the discussion.  

	The second term in the Lagrangian density,
\beqa
\label{auxact}
{\cal L}_{aux} & = &
  \p_\lambda \bar\phi_\mu^{ab} (D_\lambda \phi_\mu)^{ab}
  \\ \nonumber  
&&   - \p_\lambda \bar\omega_\mu^{ab} 
[ \ (D_\lambda \omega_\mu)^{ab} + (g D_\lambda c \times \phi_\mu)^{ab} \ ],
\eeqa
involves a quartet of auxiliary bose and fermi ghosts, $\phi_\mu^{ab}$ and $\omega_\mu^{ab}$ and corresponding anti-ghosts, $\bar\phi_\mu^{ab}$ and $\bar\omega_\mu^{ab}$ that carry a Lorentz index $\mu$ and a pair of color indices $a$ and $b$.  
The gauge-covariant derivative and the Lie commutator act on the first color index only, while the second color index is mute, thus 
$(D_\lambda \phi_\mu)^{ab} = \p_\lambda \phi_\mu^{ab} + g (A_\lambda \times \phi_\mu)^{ab}$
where $(A_\lambda \times \phi_\mu)^{ab} \equiv f^{adc}A_\lambda^d \phi_\mu^{cb}$.  

	The last term in the action involves the Gribov mass~$\gamma$, with engineering dimension~$m^4$,
\beq
\label{Lgamma}
{\cal L}_\gamma =  
 \gamma^{1/2} \ 
[ \  D_\lambda(\phi_\lambda - \bar\phi_\lambda)
-  (g D_\lambda c \times \bar\omega_\lambda) \ ]^{aa}
 - f \gamma,
\eeq 
where $f \equiv D(N^2 -1)$ is the number of components of the gluon field $A_\mu^a$.  If this term were absent, the integral over the auxiliary bose and fermi ghosts would produce cancelling factors of the Faddeev-Popov determinant $\det {\cal M}$, leaving the Faddeev-Popov action.  The term ${\cal L}_\gamma$ is of dimension 2, whereas all other terms in ${\cal L}$ are of dimension~4.  The system just defined has been studied and its renormalizability established by considering the symmetries of the 4-dimensional action~${\cal L}_0$, and treating the 2-dimensional term~${\cal L}_\gamma$ as a soft breaking of these symmetries \cite{Zwanziger:1993, Schaden:1994, Sorella:2005, Gracey:0510, Gracey:0605, Dudal:0806, Dudal:0808, Sorella:0904, Dudal:0908, Gracey:0909, Vandersickel0910, Dudal:0911, Baulieu:2009, Dudal:1001}.  

 	The Gribov mass, $\gamma$, is not a new, free parameter in QCD, which would be unacceptable, but is determined in terms of $\Lambda_{QCD}$ by the gap equation,
\beq
\label{HC}
{\p \Gamma(\gamma) \over \p \gamma} = 0,
\eeq  
where $\Gamma(\gamma)$ is the vacuum free energy, 
\beq
\label{freeenergy1}
\exp[- \Gamma(\gamma)] = \int d\Phi \ \exp[ - S(\Phi, \gamma)],
\eeq
and $\Phi_\alpha = (A, c, \bar c, b, \phi, \omega, \bar\omega, \bar\phi)$ is the set of all fields.  This gap equation is called ``the horizon condition" because it was derived from the condition that the functional integral be cut off at the Gribov horizon.  The gap equation has no solution at tree level, because
\beq
{\Gamma^{(0)}(\gamma)  \over \Omega f} = {S(\Phi, \gamma)|_{\Phi = 0}  \over \Omega f} = - \gamma,
\eeq
where~$\Omega$ is the Euclidean quantization volume, and the equation ${\p \Gamma^{(0)} \over \p \gamma} = - \Omega f = 0$ indeed has no solution.  At one-loop, the gap equation does have a solution, so it is an inherently non-perturbative condition, and yet, remarkably, it is compatible with perturbative renormalization. The one-loop contribution to the vacuum free energy,~$\Gamma^{(1)}(\gamma)$, is calculated in Appendix B, with the result
\beq
\label{gamma1loop}
{\Gamma(\gamma)  \over \Omega f} = {S(0, \gamma) + \Gamma^{(1)}(\gamma)  \over \Omega f} = - \gamma \Big[ 1
+ {  3 N g^2   \over 8 (4\pi)^2  } \ln\Big( {2N g^2 \gamma \over \mu^4} \Big) \Big],
\eeq
where $\mu$ is a renormalization mass, and $g = g(\mu/\Lambda_{QCD})$ is the renormalized running coupling constant.  For $\gamma > 0$, this function has a single stationary point\footnote{This is true for all Euclidean dimension $1 \leq D < 4$.  See Appendix B.} at \cite{Gribov:1977wm, Zwanziger:1989mf}
\beq
0 = { 1 \over \Omega f} {\p \Gamma \over \p \gamma} = - 1  - \Big( {3N g^2 \over 8 (4 \pi)^2 } \Big)  \ln \Big( { 2 N g^2 e \gamma \over \mu^4} \Big).
\eeq
This gap equation has been calculated to two-loop order~\cite{Gracey:0510, Gracey:0605}.  The last equation reads
\beq
\label{gapequation}
0 = { 1 \over \Omega f} {\p \Gamma \over \p \gamma} = { - 3 N g^2 \over 128 \pi^2} \ln \Big( {\gamma \over \gamma_{ph} } \Big)
\eeq
where 
\beq
\label{gammaphys}
\gamma_{ph} = \mu^4 { 1 \over 2 N g^2 e } \exp\Big( { - 128 \pi^2 \over 3Ng^2} \Big),
\eeq
and has the solution
\beq
\gamma = \gamma_{ph}.
\eeq

	We define a BRST operator that acts on the Faddeev-Popov fields in the usual way
\beqa
\label{BRST1}
s A_\mu & = & D_\mu c; \ \ \ \ \ \ \  \ \ \ \ \  sc = - (g/2) (c \times c) 
\nonumber  \\
s \bar c & = & ib; \ \ \ \ \ \ \ \  \ \ \ \ \ \ \    s b = 0. 
\eeqa
and that acts on the auxiliary ghosts according to
\beqa
\label{BRST2}
s \phi_\mu^{ab} & = & \omega_\mu^{ab}; 
\ \ \ \ \ \ \ \ \ \ \ \ \ \ \ \ \ \ \ \ \ \ \ \ \   s \omega_\mu^{ab} = 0
\nonumber  \\
s \bar\omega_\mu^{ab} & = & \bar\phi_\mu^{ab}; 
\ \ \ \ \ \ \ \ \ \ \ \ \ \ \ \ \ \ \ \ \ \ \ \ \   s \bar\phi_\mu^{ab} = 0.
\eeqa	
It is nil-potent, $s^2 = 0$.  This operator is a symmetry of the local lagrangian at $\gamma = 0$,
\beq
s {\cal L}_0 = 0.
\eeq  
Indeed ${\cal L}_0$ may be written in the standard form for BRST gauge fixing,
\beq
{\cal L}_0 = (1/4) F_{\mu \nu}^2 + s( \p_\mu \bar c A_\mu + \p_\lambda \bar\omega_\mu D_\lambda \phi_\mu ),
\eeq
so $s {\cal L}_0 = 0$ follows from $s F^2 = 0$ and $s^2 = 0$.  For $\gamma \neq 0$, this symmetry is explicitly but softly broken by the term~${\cal L}_\gamma$ of mass dimension 2,
\beq
\label{sonL}
s {\cal L}_\gamma =  \gamma^{1/2} [D_\lambda \omega_\lambda + g (D_\lambda c) \times \phi_\lambda]^{aa}.
\eeq

\section{The symmetry of $\Gamma$ is the symmetry of $S_0$}
\label{symGammasymS0}

	It is helpful to consider the theory just defined as a special case of a more symmetric theory.  As a first step, we write the multiplet of auxiliary ghosts as $\phi_i^a, \omega_i^a, \bar\omega^{ai}, \bar\phi^{ai}$, instead of $\phi_\mu^{ab}, \omega_\mu^{ab}, \bar\omega_\mu^{ab}, \bar\phi_\mu^{ab}$, so the index~$i$ on the auxiliary ghosts substitutes for the previous pair of indices $i = (b, \mu)$, where $b$ is the second color index, and $\mu$ the Lorentz index.  The Lagrangian density ${\cal L}_{aux}$ reads
\beqa
{\cal L}_{aux} & = &
 s  \p_\lambda \bar\omega^{ai} (D_\lambda \phi_i)^a
\nonumber \\
& = &  \p_\lambda \bar\phi^{ai} (D_\lambda \phi_i)^a
  \\ \nonumber  
&&   - \p_\lambda \bar\omega^{ai} 
[ \ (D_\lambda \omega_i)^a + (g D_\lambda c \times \phi_i)^a \ ],
\eeqa
and the BRST operator acts according to
\beqa
\label{BRST3}
s \phi_i^a & = & \omega_i^a; 
\ \ \ \ \ \ \ \ \ \ \ \ \ \ \ \ \ \ \ \ \ \ \ \ \   s \omega_i^a = 0
\nonumber  \\
s \bar\omega^{ai} & = & \bar\phi^{ai};
\ \ \ \ \ \ \ \ \ \ \ \ \ \ \ \ \ \ \ \ \ \ \ \   s \bar\phi^{ai} = 0.
\eeqa
The mute index $i$ takes on the values $i = 1 ... f$, where $f = (N^2 -1) D$, which is the number of components of the gluon field $A_\mu^a$.

	As a second step, we define the local extended action~\cite{Zwanziger:1993}.
\beqa
\label{extendaction}
\Sigma(\Phi, Q) & \equiv & S_0(\Phi) + (K_\mu, sA_\mu) + (L, sc) + (M_\lambda^i, D_\lambda \phi_i) 
 \nonumber  \\
&& + (D_\lambda \bar\omega^i, N_{\lambda i}) + (U_\lambda^i, s D_\lambda \phi_i) 
  \\  \nonumber
& &+ (s D_\lambda \bar\omega^i, V_{\lambda i}) + (M_\lambda^i, V_{\lambda i}) - (U_\lambda^i, N_{\lambda i}),
\eeqa
that depends upon local sources $Q \equiv (K_\mu^a, L^a, M_\mu^{ai}, N_{\mu i}^a, U_\mu^{ai}, V_{\mu i}^a)$ of composite operators, where $(M_\lambda^i, D_\lambda \phi_i) \equiv \int d^Dx \ M_\lambda^{a i}(x) (D_\lambda \phi_i)^a(x)$, etc.  The local sources  $K$ and $L$ are familiar from Faddeev-Popov theory.

	The action, $S_0(\Phi) = \int d^Dx \ {\cal L}_0$, appears in the definition of $\Sigma$ instead of the full action, $S(\Phi, \gamma) = \int d^Dx \ ({\cal L}_0 + {\cal L}_\gamma) = S_0 + S_\gamma$, defined in~(\ref{fullaction}).  However the full action is recovered from the extended action\beq
\label{SfromSigma}
S(\Phi, \gamma) = \Sigma(\Phi, Q_1),
\eeq
by setting the external sources to the particular values
\beqa
\label{Qone}
Q_1 \equiv \{ M_\lambda^{ai} = M_{\lambda \mu}^{ab} = - V_{\lambda, i}^a = - V_{\lambda \mu}^{ab} = \gamma^{1/2} \ \delta_{\lambda \mu} \delta^{ab}; 
\nonumber  \\
 K = L = N = U = 0\}, \ \ \ \ \ \ \ \ \ \ \ \ \    
\eeqa
where we have reverted to the previous notation, $i = (b, \mu)$.  The last two terms in the extended action $(M, V) - (U, N)$, depend only on the sources, and at $Q = Q_1$ the term $(M, V)$ takes the constant value~$(M, V)|_{Q_1} = - f \gamma$ which is the last term of ${\cal L}_\gamma$, eq. (\ref{Lgamma}).  The term $(U, N)$, which vanishes at $Q = Q_1$ increases the symmetry of $\Sigma$, as we shall see shortly.

	The extended action $\Sigma(\Phi, Q)$ has all the symmetries of the action $S_0(\Phi)$, provided that the sources $Q$ are suitably transformed.  This is a far richer set than the symmetries of the action $S$.  It includes  the Slavnov-Taylor identity that follows from the $s$-invariance of $S_0$, a $U(f)$ symmetry that acts on the index $i$, and additional symmetries found by Maggiore and Schaden~\cite{Schaden:1994} which we will turn to shortly.
	
	We repeat for emphasis.  The full action action~$S(\Phi, \gamma)$ breaks the symmetries of~$S_0(\Phi)$ softly but explicitly.  By introducing external sources $Q$, we have replaced~$S(\Phi, \gamma)$ by the extended action~$\Sigma(\Phi, Q)$ that respects all the symmetries of $S_0(\Phi)$ when the sources $Q$ are suitably transformed.  The action $S(\Phi, \gamma) = \Sigma(\Phi, Q_1)$ is recovered at a point $\Phi = 0, Q = Q_1$ that breaks the symmetries of $\Sigma(\Phi, Q)$.  We shall see shortly that this symmetry-breaking point is spontaneously chosen by the horizon condition.  
	
	To complete this section we introduce the quantum effective action $\Gamma(\Phi, Q)$, that possess the symmetries of $\Sigma(\Phi, Q)$, and thus of $S_0$.  The partition function
\beq
Z(J, Q) \equiv \int d \Phi \exp[ - \Sigma(\Phi, Q) + (J, \Phi)],
\eeq	
depends on the sources $J_\alpha$ of all elementary fields $\Phi_\alpha$ and on the sources $Q$ of composite fields.    The free energy is defined by
\beq
W(J, Q) \equiv \ln Z(J, Q),
\eeq
and the quantum effective action is obtained by Legendre transformation from $W(J, Q)$, 
\beq
\Gamma(\Phi, Q) = (J_\alpha, \Phi_\alpha) - W(J, Q)
\eeq
 where the ``classical" fields are given by
\beq
\Phi_\alpha = {\delta W \over \delta J_\alpha},
\eeq
and
\beq
\label{legendre2}
{\delta \Gamma \over \delta \Phi_\alpha} = \eta_\alpha J_\alpha.
\eeq
\beq
\label{partialQ}
{\delta \Gamma \over \delta Q_\alpha} = - {\delta W \over \delta Q_\alpha},
\eeq    
where $\eta_\alpha$ is a sign factor that depends on whether $\Phi_\alpha$ is bosonic or fermionic.  Here and below we always take the left fermionic derivative.

\section{Symmetry-breaking vacuum}

	Having generalized the parameter $\gamma$ to the local sources $M(x), V(x), U(x), N(x)$, our next step will be to express the horizon condition, ${\p \Gamma(\gamma) \over \p \gamma} = 0$, in terms of these sources.  
	
	The only restriction on our statement of the horizon condition in this more general situation is that it reduce to ${\p \Gamma(\gamma) \over \p \gamma} = 0$ at $Q = Q_1$, where $Q_1$ is defined in (\ref{Qone}).  For other values of $Q$, we may write any condition we wish, provided we reject any solution (if any there are) besides the one of the form $Q = Q_1$.  At $Q = Q_1$, only the sources $M$ and $V$ depend on $\gamma$, and at $Q = Q_1$ we write $M_{\mu \nu}^{ab}(x, \gamma)$ and $V_{\mu \nu}^{ab}(x, \gamma)$.  We have
\beqa
{\p \Gamma(M(\gamma), V(\gamma)) \over \p \gamma} = \int d^D x \ \Big( {\delta \Gamma(M, V) \over \delta M_{\mu \nu}^{ab}(x)} {\p M_{\mu \nu}^{ab}(x, \gamma) \over \p \gamma}
\nonumber \\ 
+ {\delta \Gamma(M, V) \over \delta V_{\mu \nu}^{ab}(x)} {\p V_{\mu \nu}^{ab}(x, \gamma) \over \p \gamma} \Big),
\eeqa
where all other fields and sources are set to 0.  At $Q = Q_1$,  we have, by (\ref{Qone}),
\beq
{\p M_{\mu \nu}^{ab}(x, \gamma) \over \p \gamma^{1/2} } = - {\p V_{\mu \nu}^{ab}(x, \gamma) \over \p \gamma^{1/2}  } = \delta_{\mu \nu} \delta^{ab}.
\eeq 
This yields, for the horizon condition ${\p \Gamma(\gamma) \over \p \gamma} = 0$, 
\beq
\label{HCs}
0 = \int d^D x \ \delta_{\mu \nu} \delta^{ab} \Big( {\delta \Gamma \over \delta M_{\mu \nu}^{ab}(x)}
- {\delta \Gamma \over \delta V_{\mu \nu}^{ab}(x)} \Big)\Big|_{\Phi = U = N = K = L = 0}.
\eeq	
	
 	It is shown in Appendix C that the two terms in in the last equation are equal at $Q_1$,
\beqa
\label{2termsequal}
 \int d^Dx \ \delta_{\mu \nu} \delta^{ab} { \delta \Gamma \over \delta {M}_{\mu \nu}^{ab} }\Big|_{\Phi = 0, Q = Q_1} \ \ \ \ \ \ \ \ \ \ \ \ \ \ \ \ \ 
\nonumber  \\
= - \int d^Dx \ \delta_{\mu \nu} \delta^{ab} { \delta \Gamma \over \delta {V}_{\mu \nu}^{ab} }\Big|_{\Phi = 0, Q = Q_1},
\eeqa
so, using the freedom to impose arbitrary conditions away from $Q_1$, we impose both conditions,
\beq
\label{horizona}
\int d^Dx \ \delta_{\mu \nu} \delta^{ab} { \delta \Gamma \over \delta M_{\mu \nu}^{ab} }\Big|_{\Phi = U = N = K = L = 0} = 0
\eeq
and
\beq
\label{horizonb}
\int d^Dx \ \delta_{\mu \nu} \delta^{ab} { \delta \Gamma \over \delta V_{\mu \nu}^{ab} }\Big|_{\Phi = U = N = K = L = 0} = 0.
\eeq
Moreover at $Q = Q_1$, space-time invariance, Lorentz, and global gauge invariance are respected, so we may replace the last equations by the more stringent conditions, 
\beq
\label{horizonD}
 { \delta \Gamma \over \delta {M}_{\lambda \mu}^{ab} }\Big|_{\Phi = U = N = K = L = 0} = { \delta \Gamma \over \delta {V}_{\lambda \mu}^{ab} }\Big|_{\Phi = U = N = K = L = 0} = 0.
\eeq	 
Finally we note that since $N$ and $U$ are both fermionic, we may give a more symmetric expression to these conditions by also getting $U = N = 0$ as a consequence of two more stationary conditions, so we have altogether
\beqa
\label{horizonE}
{ \delta \Gamma \over \delta M_{\lambda \mu}^{ab} }\Big|_{\Phi = K = L = 0} & = & { \delta \Gamma \over \delta V_{\lambda \mu}^{ab} }\Big|_{\Phi = K = L = 0} = 0, 
\nonumber  \\
 { \delta \Gamma \over \delta N_{\lambda \mu}^{ab} }\Big|_{\Phi = K = L = 0} & = & { \delta \Gamma \over \delta U_{\lambda \mu}^{ab} }\Big|_{\Phi = K = L = 0} = 0.
\eeqa
(We could also derive the condition $\Phi = 0$ from the stationary condition ${\delta \Gamma \over \delta \Phi_\alpha} = J_\alpha = 0$, that follows from the Legendre transformation and the absence of sources $J_\alpha$.)

	The only solution to these equations that is of interest to us is of the form
\beqa
\label{Qphys}
Q_{ph} \equiv \{ M_\lambda^{ai} = M_{\lambda \mu}^{ab} = - V_{\lambda, i}^a = - V_{\lambda \mu}^{ab} = \gamma_{ph}^{1/2} \ \delta_{\lambda \mu} \delta^{ab}; 
\nonumber  \\
 K = L = N = U = 0\}, \ \ \ \ \ \ \ \ \ \ \ \ \ \ 
\eeqa
where $\gamma_{ph}$ has a definite value, as in the one-loop calculation given above.  If there are other solutions they are rejected.

	Observe that the classical vacuum, $\Phi = 0, Q = Q_{ph}$, that is the solution to the horizon condition (\ref{horizonE}), has less symmetry than the quantum effective action $\Gamma(\Phi, Q)$.  In particular the $U(f)$ symmetry on the index $i = (b, \mu)$, noted above, is broken and, as we shall see, so are other symmetries.  Thus there are flat directions of $\Gamma$ at the classical vacuum, $\Phi = 0, Q = Q_{ph}$, that correspond to Goldstone particles.  We shall see that Goldstone bosons and Goldstone fermions both occur.  
	
	To identify the Goldstone particles and evaluate their propagators, we must determine the Ward identities and the symmetries of $\Gamma(\Phi, Q)$, and the pattern of symmetry breaking at the symmetry-breaking vacuum $\Phi = 0, Q = Q_{ph}$.  This will occupy the bulk of the present article.

\section{Slavnov-Taylor identity}

	The Slavnov-Taylor identity that results from the explicit but soft breaking of the BRST operator~$s$ has been presented before \cite{Zwanziger:1993}.  We shall rederive it here by a somewhat simpler method with a different separation of terms.

	We first derive the Slavnov-Taylor (ST) identity satisfied by~$\Sigma$.  We  use $sS_0 = 0$, which gives
\beqa
\label{softsource}
s \Sigma & = &  ( M_{\lambda \mu}, sD_\lambda \phi_\mu) + ( s D_\lambda \bar\omega_\mu, N_{\lambda \mu} )
\nonumber  \\
& = & \Big( M, {\delta \Sigma \over \delta U} + N \Big) + \Big( N, {\delta \Sigma \over \delta V} - M \Big)
\nonumber  \\
& = & \Big( M, {\delta \Sigma \over \delta U} \Big) + \Big( N, {\delta \Sigma \over \delta V} \Big).
\eeqa
The action of $s$ on $\Sigma$ may be expressed in terms of the sources,
\beqa
\label{softbreak}
s \Sigma  & \equiv &
 \int d^D x \ \Big( {\delta \Sigma \over \delta K_\mu }{\delta \Sigma \over \delta A_\mu } + {\delta \Sigma \over \delta L }{\delta \Sigma \over \delta c } + ib {\delta \Sigma \over \delta \bar c }
\nonumber    \\
&& + \omega_\mu^{ab} {\delta \Sigma \over \delta \phi_\mu^{ab} } + \bar\phi_\mu^{ab} {\delta \Sigma \over \delta \bar\omega_\mu^{ab} } \Big).
\eeqa
The last two equations yield the ST identity satisfied by the local extended action,
\beq
{\cal S}(\Sigma) = 0,
\eeq
where
\beqa
\label{STSigma}
{\cal S}(\Sigma) & \equiv &
 \Big( {\delta \Sigma \over \delta K_\mu }, {\delta \Sigma \over \delta A_\mu } \Big) + \Big( {\delta \Sigma \over \delta L }, {\delta \Sigma \over \delta c } \Big) + \Big( ib,  {\delta \Sigma \over \delta \bar c } \Big)
  \nonumber  \\
&&\ \ \  + \Big( \omega_\mu, {\delta \Sigma \over \delta \phi_\mu } \Big) + \Big( \bar\phi_\mu, {\delta \Sigma \over \delta \bar\omega_\mu } \Big)
\nonumber  \\
&& \ \ \   - \Big(  M_{\lambda \mu}, {\delta \Sigma \over \delta U_{\lambda \mu}^{ab} } \Big)  - \Big( N_{\lambda \mu}, {\delta \Sigma \over \delta V_{\lambda \mu} } \Big).
\eeqa

	The Slavnov-Taylor identity for $Z$ follows from the fact that the operator~$s$ is a derivative, and the integral of a derivative vanishes,
\beqa
\label{changevar}
0 & = & \int d \Phi \ s  \exp[ - \Sigma + (J, \Phi)]
\\    \nonumber 
& = & \int d \Phi \ [ - s \Sigma + \eta_\alpha (J_\alpha, s \Phi_\alpha) ]  \exp[ - \Sigma + (J, \Phi)],
\eeqa
where $\eta_\alpha = \pm 1$ is a sign factor that depends on whether $J_\alpha$ is bosonic or fermionic.  We use (\ref{softsource}) for $s \Sigma$, which comes outside the integral,
\beqa
\int d\Phi \ s \Sigma \  ... & = & \int d\Phi \ \Big[ \Big( M, {\delta \Sigma \over \delta U} \Big) + \Big( N, {\delta \Sigma \over \delta V} \Big) \Big] ...
\nonumber  \\
& = & - \Big[ \Big( M, {\delta \over \delta U} \Big) + \Big( N, {\delta \over \delta V} \Big) \  \Big] Z.
\eeqa 
Likewise we have
\beqa
\int d \Phi \ \eta_i (J_i, s \Phi_i) ... & = & \Big[  - \Big( J_A, {\delta \over \delta K } \Big) + \Big( J_c, {\delta \over \delta L } \Big) 
\nonumber \\ 
&& -  i\Big( J_{\bar c}, {\delta \over \delta J_b } \Big) + \Big( J_\phi, {\delta \over \delta J_\omega} \Big)
\nonumber  \\
&&  - \Big( J_{\bar\omega} {\delta \over \delta J_{\bar\phi} } \Big) \Big] Z.
\eeqa
We  thus obtain the Slavnov-Taylor identity satisfied by the partition function,
\beq
\Xi Z = 0,
\eeq
where $\Xi$ is the linear differential operator
\beqa
\Xi & = &  - \Big( J_A, {\delta \over \delta K } \Big) + \Big( J_c, {\delta \over \delta L } \Big) 
-  i\Big( J_{\bar c}, {\delta \over \delta J_b } \Big) 
+ \Big( J_\phi, {\delta \over \delta J_\omega} \Big)
\nonumber  \\
&&  - \Big( J_{\bar\omega} {\delta \over \delta J_{\bar\phi} } \Big) + \Big( M, {\delta \over \delta U} \Big) + \Big( N, {\delta \over \delta V} \Big).
\eeqa

	The free energy $W(J, Q) \equiv \ln Z(J, Q)$
satisfies the same equation,
\beqa
\label{STW}
&&  - \Big( J_A, {\delta W \over \delta K } \Big) + \Big( J_c, {\delta W \over \delta L } \Big) 
-  i\Big( J_{\bar c}, {\delta W \over \delta J_b } \Big) 
+ \Big( J_\phi, {\delta W \over \delta J_\omega} \Big)
\nonumber  \\
 && - \Big( J_{\bar\omega} {\delta W \over \delta J_{\bar\phi} } \Big) + \Big( M, {\delta W \over \delta U} \Big) + \Big( N, {\delta W \over \delta V} \Big) = 0.
\eeqa

	It follows from the Legendre transformation~(\ref{legendre2}) and~(\ref{partialQ}) that the quantum effective action~$\Gamma$ satisfies the same Slavnov-Taylor identity as the local action~$\Sigma$,
\beq
\label{STx}
{\cal S}(\Gamma) = 0,
\eeq
where ${\cal S}(\Gamma)$ is defined in (\ref{STSigma}).

\section{Reduced Quantum Effective Action}

	In Appendix A we solve the Ward identities that correspond to the equations of motion of the fields $\bar c, b, \phi, \omega, \bar\omega, \bar\phi$.  This gives the complete dependence of the quantum effective action $\Gamma$ on these fields \cite{Zwanziger:1993}.  Also in Appendix A, the Ward identity corresponding to the integrated equation of motion of the Faddeev-Popov ghost field $c$ is solved.  This is believed to be a new result.  
	
		According to eqs.~(\ref{primesolution}), (\ref{spsv}), (\ref{primetostar}), (\ref{reducedstar}), and~(\ref{partialc}),~$\Gamma$ is expressed in terms of the reduced quantum effective action~$\Gamma^*$ 
\beq
\label{finalsolution}
\Gamma = \Sigma_{inv} + \Gamma^*(A, \p c, k, L, m, n, u, v),
\eeq
that depends on a reduced number of variables defined by\footnote{A more consistent notation for the reduced variables would  be $M^*, N^*, U^*, V^*, K*$ instead of $m, n, u, v, k$, but this would give the equations a rather baroque appearance.}
\beqa
\label{starsources}
m_{\lambda \mu} & \equiv & M_{\lambda \mu} + \p_\lambda \bar\phi_\mu - g c \times (U_{\lambda \mu} - \p_\lambda \bar\omega_\mu)
\nonumber  \\
n_{\lambda \mu} & \equiv & N_{\lambda \mu} - \p_\lambda \omega_\mu - g c \times (V_{\lambda \mu} + \p_\lambda \phi_\mu)
\nonumber  \\
u_{\lambda \mu} & \equiv & U_{\lambda \mu} - \p_\lambda \bar\omega_\mu
\nonumber  \\
v_{\lambda \mu} & \equiv & V_{\lambda \mu} + \p_\lambda \phi_\mu
\nonumber  \\
k_\lambda^d & \equiv & K_\lambda^d - \p_\lambda \bar c^d - g (\bar\omega_\mu \times V_{\lambda \mu})^d
\nonumber  \\
&& \ \ \ \  - [ g (U_{\lambda \mu} - \p_\lambda \bar\omega_\mu) \times \phi_\mu ]^d,
\eeqa
and
\beqa
\label{Sigmainvariant}
\Sigma_{inv} \equiv (i \p_\lambda b, A_\lambda) 
 + (M_{\lambda \mu} + \p_\lambda \bar\phi_\mu, g A_\lambda \times \phi_\mu)
 \nonumber \\ 
+ (g A_\lambda \times \bar\phi_\mu, V_{\lambda \mu}) + ( g A_\lambda \times \bar\omega_\mu, N_{\lambda \mu}) 
 \nonumber \\
+ (U_{\lambda \mu} - \p_\lambda \bar\omega_\mu, g A_\lambda \times \omega_\mu) + (k_\lambda, gA_\lambda \times c)
 \nonumber \\
 + (L, (-g/2)(c \times c)).
\eeqa
This gives the complete dependence of $\Gamma$ on $b, \bar c$ and on the 4 auxiliary ghosts $\phi, \omega, \bar\omega, \bar\phi$.  Moreover $\Gamma^*$ depends only on the derivatives $\p_\mu c$ of $c$, but not on $c$ itself.

	Each term in $\Sigma_{inv}$ contains at least one ghost field $(c, \phi, \bar\phi, \omega, \bar\omega)$ that is not differentiated whereas, according to (\ref{starsources}), only the derivatives of these ghost fields appear in the reduced action $\Gamma^*$, but not the ghost fields themselves.  Consequently $\Gamma^*$ contains no radiative corrections to the terms in $\Sigma_{inv}$, and they are invariant under renormalization.  This includes in particular the gluon-ghost mixing term $(M_\lambda^i, g(A_\lambda \times \phi_i)) + ( g(A_\lambda \times \bar\phi^i), V_{\lambda i} )$ which has the value
 $\gamma^{1/2} g  f^{abc} A_\lambda^b (\phi - \bar\phi)_\lambda^{ca}$ at the values of the sources $Q = Q_1$.
	
	The local extended action $\Sigma$ has the same decomposition as $\Gamma$,
\beq
\Sigma = \Sigma_{inv} + \Sigma^*(A, \p c, k, L, m, n, u, v),
\eeq	  
where the reduced local action is given by
\beq
\label{reducedlocal}
\Sigma^* = \int d^Dx \  [ (1/4) F_{\mu \nu}^2 + k_\lambda \p_\lambda c
+ m_{\lambda \mu} v_{\lambda \mu} - u_{\lambda \mu} n_{\lambda \mu} ].
\eeq	
When the sources $Q$ are given the values $Q_1$, eq.~(\ref{Qone}), that correspond to the action $S(\Phi, \gamma) = \Sigma(\Phi, Q_1)$,  one has
\beqa
\label{Sigmainvariant1}
\Sigma_{inv}(Q_1) & = & (i \p_\lambda b, A_\lambda) 
 + (\gamma^{1/2} \delta_{\lambda \mu}I + \p_\lambda \bar\phi_\mu, g A_\lambda \times \phi_\mu)
 \nonumber \\ 
&& - \gamma^{1/2} (g A_\lambda \times \bar\phi_\mu, \delta_{\lambda \mu} I) 
- ( \p_\lambda \bar\omega_\mu, g A_\lambda \times \omega_\mu)
\nonumber \\
&&+ (k_\lambda, g A_\lambda \times c) + (L, (-g/2) c \times c), 
\eeqa	
where $I^{ab} = \delta_{ab}$.  Thus $\Sigma_{inv}(Q_1)$ contains the gauge-fixing term, $(i \p_\lambda b, A_\lambda)$, all ghost-ghost-gluon vertices, and the quadratic terms $\int d^D x \ \gamma^{1/2} g f^{abc} A_\lambda^b (\phi - \bar\phi)_\lambda^{ca}$ that are responsible for gluon-ghost mixing.  These terms are all invariant under renormalization.

	Expressions (\ref{finalsolution}) and (\ref{starsources}) for the quantum effective action~$\Gamma$ severely restrict possible counter terms.  For example, no mass terms such as $m^2 \bar\phi_\mu \phi_\mu$ are allowed.

\section{Slavnov-Taylor identity satisfied by reduced quantum effective action}

	We shall derive the Slavnov-Taylor identity satisfied by the reduced quantum effective action in two steps.
	
	{\it Step 1.}  We substitute expression (\ref{primesolution}) for $\Gamma$ into the ST identity (\ref{STx}) to obtain an identity satisfied by the partially reduced quantum effective action $\Gamma'$.  We first rewrite (\ref{STx}) as
\beq
\label{STy}
{\cal S}(\Gamma) = 
\Big( { \delta \Gamma \over \delta K}, { \delta \Gamma \over \delta A} \Big) + \Big( { \delta \Gamma \over \delta L}, { \delta \Gamma \over \delta c} \Big) + {\cal L} \Gamma = 0,
\eeq	
where we have separated out the part that is linear in $\Gamma$,
\beqa
\label{linearinG}
{\cal L} \Gamma & \equiv & \Big[  \Big( ib,  {\delta \over \delta \bar c } \Big)
  + \Big( \omega_\mu, {\delta \over \delta \phi_\mu } \Big) + \Big( \bar\phi_\mu, {\delta \over \delta \bar\omega_\mu } \Big)
\nonumber  \\
&& \ \ \   - \Big(  M_{\lambda \mu}, {\delta \over \delta U_{\lambda \mu}^{ab} } \Big)  - \Big( N_{\lambda \mu}, {\delta \over \delta V_{\lambda \mu} } \Big) \Big] \Gamma,
\eeqa
and we find
\beq
{\delta \Sigma_p \over \delta K} = {\delta \Sigma_p \over \delta c} = {\delta \Sigma_p \over \delta L} = {\cal L} \Sigma_p = 0,
\eeq
and
\beq
{\delta \Sigma_p \over \delta A_\lambda} = - Y_\lambda,
\eeq
where
\beqa
Y_\lambda \equiv - i \p_\lambda b + g(U_{\lambda \mu} - \p_\lambda \bar\omega_\mu) \times \omega_\mu - g\bar\omega_\mu \times N_{\lambda \mu}
\nonumber  \\
 + g M'_{\lambda \mu} \times \phi_\mu - g\bar\phi_\mu \times (V_{\lambda \mu} - \gamma^{1/2} \delta_{\lambda \mu} I),
\eeqa
so the contribution to ${\cal S}(\Gamma)$ from $\Sigma_p$ is given by
\beq
 - \Big( { \delta \Gamma' \over \delta K}, Y \Big).
\eeq
We also evaluate the partial derivatives,
\beq
{\delta \Gamma' \over \delta \bar c} = \p_\lambda {\delta \Gamma' \over \delta K'_\lambda}
\eeq
\beq
{\delta \Gamma' \over \delta U_{\lambda \mu} } = {\delta \Gamma' \over \delta U'_{\lambda \mu} } - g \phi_\mu \times{\delta \Gamma' \over \delta K'_\lambda }
\eeq
\beq
{\delta \Gamma' \over \delta V_{\lambda \mu} } = {\delta \Gamma' \over \delta V'_{\lambda \mu} } + g \bar\omega_\mu \times{\delta \Gamma' \over \delta K'_\lambda }
\eeq
\beq
{\delta \Gamma' \over \delta \phi_\mu } = - \p_\lambda {\delta \Gamma' \over \delta V'_{\lambda \mu} } + g(U_{\lambda \mu} - \p_\lambda\bar\omega_\mu) \times{\delta \Gamma' \over \delta K'_\lambda }
\eeq
\beqa
{\delta \Gamma' \over \delta \bar\omega_\mu } =  \p_\lambda {\delta \Gamma' \over \delta U'_{\lambda \mu} } - \p_\lambda \Big( g \phi_\mu \times {\delta \Gamma' \over \delta K'_\lambda } \Big)
\nonumber  \\
 - g (V_{\lambda \mu} - \gamma^{1/2} \delta_{\lambda \mu} I) \times  {\delta \Gamma' \over \delta K'_\lambda },
\eeqa
which yields
\beq
{\cal L} \Gamma' = - \Big( M'_{\lambda \mu}, {\delta \Gamma'  \over \delta U'_{\lambda \mu} } \Big)
- \Big( N'_{\lambda \mu}, {\delta \Gamma' \over \delta V'_{\lambda \mu} } \Big) +  \Big( Y_\lambda, {\delta \Gamma' \over \delta K'_\lambda } \Big).
\eeq 
The terms in $Y_\lambda$ cancel, and the Slavnov-Taylor identity simplifies to
\beqa
\label{STq}
{\cal S}'(\Gamma') & \equiv & \Big( { \delta \Gamma' \over \delta K'}, { \delta \Gamma' \over \delta A} \Big) + \Big( { \delta \Gamma' \over \delta L}, { \delta \Gamma' \over \delta c} \Big)  
\nonumber  \\
&& - \Big( M'_{\lambda \mu}, {\delta \Gamma' \over \delta U'_{\lambda \mu} } \Big)
- \Big( N'_{\lambda \mu}, {\delta \Gamma'  \over \delta V'_{\lambda \mu} } \Big)  = 0.
\eeqa

	{\it Step 2.}  We make the change of variable (\ref{primetostar}) and (\ref{reducedstar}) with the result that the reduced quantum effective action satisfies the Slavnov-Taylor identity
\beq
\label{STreduced}
{\cal S}^*(\Gamma^*) \equiv {\cal S}_0^*(\Gamma^*) + ({\cal L}_1^* + {\cal L}_2^*) \Gamma^* = 0,
\eeq
where
${\cal S}_0^*(\Gamma^*)$ is bilinear in $\Gamma^*$,
\beqa
{\cal S}_0^*(\Gamma^*) & \equiv & \Big( { \delta \Gamma^* \over \delta k_\lambda}, { \delta \Gamma^* \over \delta A_\lambda} \Big) 
\\  \nonumber
&& + \Big( { \delta \Gamma^* \over \delta L}, { \delta \Gamma^* \over \delta c } 
- g u_\lambda \times { \delta \Gamma^* \over \delta m_\lambda} - g v_\lambda \times { \delta \Gamma^* \over \delta n_\lambda} \Big),
\eeqa
and 
\beqa
{\cal L}_1^* \equiv - \Big( m, { \delta \over \delta u } \Big) -  \Big( n, { \delta \over \delta v } \Big) 
\nonumber  \\
- \Big( g k_\lambda \times A_\lambda, { \delta \over \delta L } \Big)
\eeqa
\beqa
\label{Lsubtwo}
{\cal L}^*_2 & \equiv & - g \Big( c, A_\lambda \times {\delta \over \delta A_\lambda} 
+ k_\lambda \times {\delta \over \delta k_\lambda} 
+ (1/2) \ c \times {\delta \over \delta c}
  \nonumber  \\
&& \ \ \ \ \   + L \times {\delta \over \delta L} 
+ m^i_\lambda \times {\delta \over \delta m^i_\lambda } 
+ n_{\lambda i} \times {\delta \over \delta n_{\lambda i} }
\nonumber  \\
&& \ \ \ \ \ \  + u^i_\lambda \times {\delta \over \delta u^i_\lambda }
+ v_{\lambda i} \times {\delta \over \delta v_{\lambda i} } \Big).
\eeqa

	The linearized form of ${\cal S}^*(\Gamma^*)$, defined by 
\beq
\delta {\cal S}^*(\Gamma^* ) =  B_{\Gamma^*} \delta \Gamma^*
\eeq	
is given by
\beq
\label{STqa}
B_{\Gamma^*} =  \int d^Dx \ \Big( { \delta \Gamma^* \over \delta k} { \delta  \over \delta A} +  { \delta \Gamma^* \over \delta A} { \delta  \over \delta k} + {\cal L}_1^* + {\cal L}_2^* + {\cal L}_3^* \Big).
\eeq
where
\beqa
{\cal L}_3^* \equiv \Big( { \delta \Gamma^* \over \delta L}, { \delta \over \delta c } 
- g u_\lambda \times { \delta \over \delta m_\lambda} - g v_\lambda \times { \delta \over \delta n_\lambda} \Big)
\nonumber \\
+ \Big( { \delta \Gamma^* \over \delta c } 
- g u_\lambda \times { \delta \Gamma^* \over \delta m_\lambda} - g v_\lambda \times { \delta \Gamma^* \over \delta n_\lambda}, { \delta \over \delta L} \Big).
\eeqa
This operator is nil-potent,
\beq
B_{\Gamma^*}^2 = 0,
\eeq
by virtue of the ST identity ${\cal S}^*(\Gamma^*) = 0$.

	The reduced ST identity and the linear operator $B_{\Gamma^*}$ are independent of $\gamma$ and $g$.  They expresses the geometric character of a quantum gauge theory, which is the same for $\gamma = 0$ and $\gamma \neq 0$.

\section{Global Ward identities}

	An extensive set of Ward identities for the quantum effective action $\Gamma$ is provided in \cite{Schaden:1994}.  Here we shall express them as identities satisfied by the quantum effective action $\Gamma^*$.  
		
	In Appendix A it is shown that the generator of translation of the ghost $c$ by a constant,
\beq
{\cal G}^{*a} = \int d^Dx \ {\delta \over \delta c^a}
\eeq
is a symmetry of $\Gamma^*$, ${\cal G}^{*a} \Gamma^* = 0$.  If we commute ${\cal G}^{*a}$ with~${\cal S}^*$, we get the generator of rigid gauge transformations,
\beq
{\cal G}^{*a} {\cal S}^*(\Gamma^*) - B_{\Gamma^*}{\cal G}^{*a} \Gamma^* = {\cal H}_{rig}^{*a} \Gamma^*,
\eeq
\beqa
\label{Hrigid}
{\cal H}_{rig}^{*a} & \equiv & - g \Big(A_\lambda \times {\delta \over \delta A_\lambda} 
+ k_\lambda \times {\delta \over \delta k_\lambda} 
+ c \times {\delta \over \delta c}
  \nonumber  \\
&& \ \ \ \ \   + L \times {\delta \over \delta L} 
+ m^i_\lambda \times {\delta \over \delta m^i_\lambda } 
+ n_{\lambda i} \times {\delta \over \delta n_{\lambda i} }
\nonumber  \\
&& \ \ \ \ \ \  + u^i_\lambda \times {\delta \over \delta u^i_\lambda }
+ v_{\lambda i} \times {\delta \over \delta v_{\lambda i} } \Big)^a,
\eeqa
as is obvious from (\ref{Lsubtwo}).  It is a symmetry of $\Gamma^*$,
\beq
{\cal H}_{rig}^{*a} \Gamma^* = 0.
\eeq

	The fermionic operator,
\beq
{\cal R}_i^j \equiv \int d^Dx \ \Big( \phi_i^a {\delta \over \delta \omega_j^a} - \bar\omega^{a j} {\delta \over \delta \bar\phi^{a i} } - V_{\lambda i}^a {\delta \over \delta N_{\lambda j}^a } + U_\lambda^{aj} {\delta \over \delta M_\lambda^{a i} } \Big),
\eeq	  	
is a super-symmetry of the extended local action
\beq
{\cal R}_i^j \Sigma = 0.
\eeq
Here we have used the notation $\phi_i^a \equiv \phi_\mu^{ab}$, where $i = (\mu, b)$ takes the values $i  = 1, ... f$ where $f \equiv D(N^2 -1)$, and similarly for $\bar\phi^{ja} \equiv \phi_\mu^{ab}$, where $j = (\mu, b)$, etc.  In $S_0$ the indices $i$ and $j$ are mute internal indices that carry a $U(f)$ symmetry.

	Because ${\cal R}_i^j$ acts linearly on the fields, it is also a super-symmetry of the quantum effective action	
\beq
{\cal R}_i^j \Gamma = 0.
\eeq	
We write $\Gamma = \Sigma_{inv} + \Gamma^*$, where $\Gamma^*$ is a reduced action, and make the change of variables, defined in~(\ref{finalsolution}) and~(\ref{starsources}).  We have
\beq
{\cal R}_i ^j \Sigma_{inv} = 0,
\eeq
and we obtain the reduced Ward identity
\beq
{\cal R}_i^{*j} \Gamma^* = 0,
\eeq
where
\beq
\label{Rstar}
{\cal R}_i^{*j}  \equiv \int d^Dx \ \Big(  - v_{\lambda i}^a {\delta \over \delta n_{\lambda j}^a } + u_\lambda^{aj} {\delta \over \delta m_\lambda^{ai} } \Big).
\eeq

	If we anti-commute ${\cal R}_i^{*j}$ with ${\cal S}^*$, \beq
{\cal R}_i^{*j} {\cal S}^*(\Gamma^*) + B_{\Gamma^*}{\cal R}_i^{*j} \Gamma^* = {\cal U}_i^{*j}(\Gamma^*),
\eeq
we get the generator of global U(f) transformations that acts on the $i$ and $j$ indicies,
\beqa
\label{Ustar}
{\cal U}_i^{*j}  \equiv \int d^Dx \ \Big( - m_\lambda^{aj} {\delta \over \delta m_\lambda^{ai} } + n_{\lambda i}^a {\delta \over \delta n_{\lambda j}^a }
\nonumber \\
- u_\lambda^{aj} {\delta \over \delta u_\lambda^{ai} }
  + v_{\lambda i}^a {\delta \over \delta v_{\lambda j}^a }  \Big).
\eeqa

	The bosonic operator,
\beq
{\cal F}^j \equiv \int d^Dx \ \Big( c^a {\delta \over \delta \omega_j^a} - \bar\omega^{a j} {\delta \over \delta \bar c^a } - U_\lambda^{aj} {\delta \over \delta K_\lambda^a } \Big),
\eeq	  	
is another symmetry of the extended local action
\beq
{\cal F}^j \Sigma = 0.
\eeq
It also acts linearly on the fields, and is a symmetry of the quantum effective action	
\beq
{\cal F}^j \Gamma = 0.
\eeq	
We have $\Gamma = \Sigma_{inv} + \Gamma^*$, where $\Gamma^*$ is the reduced action,  and find
\beq
{\cal F}^j \Sigma_{inv} = 0.
\eeq
Under the change of variable (\ref{finalsolution}) and (\ref{starsources}) we obtain the reduced Ward identity
\beq
{\cal F}^{*j} \Gamma^* = 0,
\eeq
where
\beq
\label{Fstar}
{\cal F}^{*j} \equiv \int d^Dx \ \Big( - \p_\lambda c^a {\delta \over \delta n_{\lambda j}^a } - u_\lambda^{aj} {\delta \over \delta k_\lambda^a } \Big).
\eeq

	If we commute ${\cal F}^{*i}$ with~${\cal S}^*$,
\beq
{\cal F}^{*j} {\cal S}^*(\Gamma^*) - B_{\Gamma^*}{\cal F}^{*j} \Gamma^* = {\cal T}^{*j}(\Gamma^*),
\eeq
we get the new functional,
\beqa
{\cal T}^{*j}(\Gamma^*) & \equiv & 
\int d^Dx \ \Big[ - { \delta \Gamma^* \over \delta L^a } \Big( \p_\lambda { \delta \Gamma^* \over \delta n_{\lambda j}^a} + g (A_\lambda \times u_\lambda^j)^a \Big)
\nonumber  \\
&& \ \ \ \ \ \ \ \ \ \ \ \ 
+ \p_\lambda c^a {\delta \Gamma^* \over \delta v_{\lambda j}^a } - m_\lambda^{aj}  {\delta \Gamma^* \over \delta k_\lambda^a } \Big], 
\eeqa
which provides a new Ward identity by virtue of the preceding identities,
\beq
{\cal T}^{*j}(\Gamma^*) = 0.
\eeq

\section{Invariants under linear symmetries}

	The U(f) symmetry is a global symmetry under which the fields transform linearly.  This symmetry is realized in~$\Gamma^*$ by summing over upper and lower indices $i$  or $j$ such as, for example, $\p_\kappa u_\lambda^{ai}  \p_\mu n_{\lambda i}^b$ or~$u_\lambda^{ai}  v_{\mu i}^b$, and $\Gamma^*$ is a function of the tensor invariant obtained by this sum over~$i$.
	
	The symmetries generated by ${\cal R}_i^{*j}$ and ${\cal F}^{*j}$ are also global symmetries under which the variables transform linearly, and we wish to form invariants under these symmetries also.  Because~${\cal R}_i^{*j}$  is a super-symmetry operator, the basic tensor invariant of which all others are constructed is obtained from contracting two super-multiplets.  From the form of ${\cal R}_i^{*j}$, given in (\ref{Rstar}), we see that $m$ and $n$ must appear together, multiplied respectively by $v$ and $u$, and from the form of ${\cal F}^{*j}$, given in (\ref{Fstar}), we see that  $n$ and $k$ must appear together, multiplied respectively by $u$ and $\p c$.    This suggests grouping these fields into the super-multiplets
\beqa
\psi_{\lambda A}^a \equiv (v_{\lambda i}^a,  n_{\lambda j}^a, \p_\lambda c^a)
\nonumber  \\
\bar\psi_\lambda^{aA} \equiv (m_\lambda^{ai},  - u_\lambda^{aj}, k_\lambda^a ),
\eeqa
which contain $f$ bosonic and $f+1$ fermionic components, and the index $A= (i, j + f, 2f+1)$ takes on $2f + 1$ values.  The requirement that quantities with derivatives, such as $\p_\kappa \bar\psi_\lambda^{aA} \p_\mu \psi_{\nu A}^b$  be ${\cal R}$- and ${\cal F}$-invariants determines that the multiplets be formed as stated and not, for example, as  $(m_\lambda^{ai},  n_{\lambda j}^a, \p_\lambda c^a)$ and $(v_{\lambda i}^a,  - u_\lambda^{aj}, k_\lambda^a )$.  The fields or sources $n, m$ and $k$ can only appear in combinations where upper and lower indices $A$ are contracted.  For example, the local extended action
\beq
\Sigma =  \Sigma_{inv} + \Sigma^*,
\eeq
is expressed by
\beq
\label{Sigmastar}
\Sigma^* = \int d^Dx \ [ (1/4) (F_{\mu \nu}^a)^2 + \bar\psi_\mu^{aA} \psi_{\mu A}^a ],
\eeq
where
\beq
 \bar\psi_\mu^{aA} \psi_{\mu A}^a = m_\mu^{ai} v_{\mu i}^a - u_\mu^{ai} n_{\mu i}^a + k_\mu^a \p_\mu c^a.
\eeq
In Faddeev-Popov theory we have $\bar\psi_\mu^{aA} \psi_{\mu A}^a =  k_\mu^a \p_\mu c^a$, and we note in passing that (\ref{Sigmastar}) shows that the present theory with auxiliary bosons has the same number of independent renormalization constants as Faddeev-Popov theory in Landau gauge, namely 2. We call invariants that are formed by contraction on the $A$-indices ``$\bar\psi \psi$-invariants".  They have fermi-ghost number~0.

	In our analysis of symmetry-breaking to find the Goldstone and non-Goldstone modes we shall be interested in possible non-derivative terms in $\Gamma^*$, for they are dominant in the infrared.  They may be formed from the basic invariant tensor
\beq
\label{tensor}
T_{\mu \nu}^{ab} \equiv  \bar\psi_\mu^{aA} \psi_{\nu A}^b. 
\eeq
Allowed non-derivative terms must also be Lorentz and global color invariant, and are formed by contraction on the Lorentz and color indices.  Possible invariant terms in $\Gamma$ are thus
\beq
\label{tensorinvariants}
T_{\mu \mu}^{aa}, \ \  (T_{\mu \mu}^{aa})^2, \ \  T_{\mu \nu}^{ab} T_{\mu \nu}^{ab}, \ \  T_{\mu \nu}^{ab}T_{\mu \nu}^{ba}, ... \ .
\eeq	

	The linear symmetries ${\cal R}^*$ and~${\cal F}^*$, leave $u$ and $v$ invariant,
\beq
{\cal R}_i^{*j} u_\lambda^{ak} = {\cal R}_i^{*j} v_{\lambda k}^a = {\cal F}^{*j} u_\lambda^{ak} = {\cal F}^{*j} v_{\lambda k}^a = 0,
\eeq	
so we may also make invariants under these symmetries and $U(f)$ by contracting the $i$ indices of $u^i$ and $v_i$, such as for example
\beq
\label{uvtype}
\p_\kappa u_\lambda^{ai} \ \p_\mu v_{\nu i}^b.
\eeq
Furthermore $c$ and $L$ are separately invariant under the linear symmetries, as is $A$, and moreover $c$ appears only as the derivative $\p_\mu c$.  There are no other invariants under the linear symmetries $U(f), {\cal R}^*$, and ${\cal F}^*$  

	We now consider the consequences of conservation of  total fermi-ghost number.  The invariant of $uv$-type, such as (\ref{uvtype}), has fermi-ghost number -1, so it appears in $\Gamma$ only in association with an invariant that has fermi-ghost number + 1.  The only invariant under the linear symmetries with fermi-ghost +1 is $\p_\mu c$.  So the $uv$-invariant must appear in the combination $uv\p c$, for example
\beq
\p_\kappa u_\lambda^{ai} \ \p_\mu v_{\nu i}^b \p_\sigma c^d.
\eeq
We call these ``$uv \p c$-invariants."  The only other invariant with fermi-ghost number 0 is of $L \p c \p c$-type, such as
\beq
L^a \p_\mu c^b  \p_\nu c^d. 
\eeq  
These are the only invariants under the linear symmetries with fermi-ghost number 0.  Since the reduced quantum effective action $\Gamma^*$ has fermi-ghost number zero, the only invariants that appear in it are of the 3 types we have found,
\beq
\label{3invariants}
\bar\psi \psi, \ \ \ \  uv \p c, \ \ \ \  L \p c \p c,
\eeq
which may be freely combined with $A$.  

	We assign separate Faddeev-Popov and auxilliary fermi-ghost number to the various fields and sources as shown.
	
\vspace{.5cm} 
\begin{center}
\begin{tabular}{c | c c c c c c c c}
  & A & c & K& L & m & n & u & v \\
  \hline
FP & 0 & 1 & -1 & -2 & 0 & 0 & 0 & 0\\
aux & 0 & 0 & 0 & 0 & 0 & 1 & -1 & 0\\
\end{tabular}
\end{center}		 
Of the 3 possible invariant types with fermi-ghost number 0, the $\bar\psi \psi$ and $L \p c \p c$ types separately conserve Faddeev-Popov and auxiliary fermi-ghost number, whereas $uv \p c$ type has Faddeev-Popov fermi-ghost number +1 and auxiliary fermi-ghost number -1.  Thus, although Faddeev-Popov and auxiliary fermi-ghost number are not separately conserved, there is no invariant with negative Faddeev-Popov ghost number.  Consequently when we decompose $\Gamma^*$ into terms with definite Faddeev-Popov fermi-ghost number $n$, no negative terms appear in the sum
\beq
\label{nonegative}
\Gamma^* = \sum_{n = 0}^\infty \Gamma^{*n}.
\eeq  
This will simplify the calculation of the propagators.

	To obtain invariants under the Slavnov-Taylor identity is more difficult because it is non-linear.  The following statement is useful in this regard, but it is not as strong as one would like because it does not solve the problem of $A$-dependence.

	{\it Statement: an action formed from linear combinations of the tensor invariants given in (\ref{tensorinvariants}) satisfies all the Ward identities and the Slavnov-Taylor identity (\ref{STreduced}) of the reduced quantum effective action $\Gamma^*$.}
	
	We shall prove this for a particular example, but the proof is easily generalized to any linear combination of the tensor invariants.  For our example we take
\beq
I \equiv \int d^D x \ (1/2) ( T_{\mu \mu}^{aa} + f \gamma_{ph})^2,
\eeq	
where $T_{\mu \mu}^{aa} \equiv \bar\psi_\mu^{a B} \psi_{\mu B}^a$.  It is straightforward to verify that the global Ward identities are satisfied,
\beq
{\cal R}_i^{*j} I = {\cal F}^{*j} I =  {\cal T}^{*j} I = 0.
\eeq	
To show that $I$ satisfies the ST identity (\ref{STreduced})
\beq
{\cal S}^*(I) = 0.
\eeq
we observe that 
\beq
{\delta I \over \delta A} = {\delta I \over \delta L} = 0,
\eeq
so ${\cal S}_0^*(I) = 0$.  We also have	
\beq
{\delta I \over \delta u^{aj} } = - (1/2) \ n_j^a (  T_{\mu \mu}^{aa}+ f \gamma_{ph}).
\eeq
\beq
{\delta I \over \delta v_j^a} = (1/2) m_a^j (  T_{\mu \mu}^{aa} + f \gamma_{ph}),
\eeq
so
\beq
\Big(m, {\delta I \over \delta u} \Big) +  \Big(n, {\delta I \over \delta v} \Big) = 0,
\eeq
and we conclude ${\cal L}_1^*I = 0$.  The only slightly non-trivial part of the calculation of ${\cal L}_2^*I$ involves  
\beqa
&& 2 \int d^Dx \ c \cdot \Big( k_\lambda \times { \delta I \over \delta k_\lambda} + (1/2) c \times 
{ \delta I \over \delta c} \Big) 
\nonumber  \\
&& =  \int d^Dx \ c \cdot \Big( ( k_\lambda \times  \p_\lambda c ) ( T_{\mu \mu}^{aa} + f \gamma_{ph})
\nonumber \\
 && \ \ \ \ \ \ \ \ \ \ \ \ \ \ \ \ \ \ \  + (1/2) ( c \times \p_\lambda [ k_\lambda ( T_{\mu \mu}^{aa} + f \gamma_{ph})] \Big)  
 \nonumber  \\
 && = \int d^Dx [ ( c \times \p_\lambda c) - (1/2) \p_\lambda (c \times c) ]  k_\lambda ( T_{\mu \mu}^{aa} + f \gamma_{ph})
 \nonumber \\
&& = 0.
\eeqa
We conclude that ${\cal L}^*_2 I = 0$, so the ST identity, ${\cal S}^*(I) = 0$, is satisfied. $\Box$

\section{On-shell symmetries}

	We shall not attempt to further exploit BRST symmetry.  Instead we shall go on-shell and consider on-shell symmetries.  The only appearance of $\bar c$ in the action $S$ occurs in the Faddeev-Popov ghost action $(\p_\mu \bar c, D_\mu c)$.  We cancel all other terms in the action that are linear in $c$, and thus of the form $(L, c)$, by the shift
\beq
\label{cbarshift}
\bar c \to {\bar c} - M^{-1} L.
\eeq
This cancels the third term in 
\beqa
\label{Saux}
S_{aux} & = &
 \int d^Dx \ \{  \p_\lambda \bar\phi^{ai} (D_\lambda \phi_i)^a
  \\ \nonumber  
&&   - \p_\lambda \bar\omega^{ai} 
[ \ (D_\lambda \omega_i)^a + (g D_\lambda c \times \phi_i)^a \ ] \ \}.
\eeqa
which simplifies to
\beq
S_{aux} =
 \int d^Dx \ [ \  \p_\lambda \bar\phi^{ai} (D_\lambda \phi_i)^a   - \p_\lambda \bar\omega^{ai} 
 \ (D_\lambda \omega_i)^a \ ].
\eeq
This allows us to ignore off-diagonal correlators of the type $\bar\omega c$, which correspond to $n \geq 1$ in (\ref{nonegative}).
	
	We form the real and imaginary parts of the bose ghost,
\beq
\label{realparts}
X_i^a \equiv (\phi_i^a + \bar\phi^{ai})/\sqrt 2; \ \ \ \ \ Y_i^a \equiv (\phi_i^a - \bar\phi^{ai})/\sqrt 2 i,
\eeq
and, after integrating by parts, we obtain
\beqa
S_{aux} & = &
 \int d^Dx \ [ \ (1/2) \p_\lambda X_i^a (D_\lambda X_i)^a
 \nonumber  \\
 && + (1/2) \p_\lambda Y_i^a (D_\lambda Y_i)^a
 - X_i^a Y_i^c g f^{abc} \p_\lambda A_\lambda^b
 \nonumber  \\
 &&  \ \ \ \ \ \ \ \ \ \ \ 
 - \p_\lambda \bar\omega^{ai} 
 \ (D_\lambda \omega_i)^a \ ].
\eeqa
We now integrate out the Lagrange multiplier field $b$, so the transversality condition $\p_\lambda A_\lambda = 0$ is satisfied identically (on-shell gauge condition).  This makes the Faddeev-Popov operator symmetric, $M = - \p_\lambda D_\lambda = - D_\lambda \p_\lambda = M^\dag$, and we obtain	
\beqa
S_{aux} & = &
 \int d^Dx \ [ \ (1/2) \p_\lambda X_i^a (D_\lambda X_i)^a
 \nonumber  \\
 &&  \ \ \ \ \ \ \ \ \ 
 + (1/2) \p_\lambda Y_i^a (D_\lambda Y_i)^a
 \nonumber  \\
 &&  \ \ \ \ \ \ \ \ \ \ \ 
 - \p_\lambda \bar\omega^{ai} 
 \ (D_\lambda \omega_i)^a \ ].
\eeqa

	To display the symmetries of the bosonic part of the action, we form the vector
\beq
Z_p^a = (X_i^a, Y_j^a),
\eeq 	
where $p = 1, . . . 2f$, so the action reads
\beq
S_{aux} =
 \int d^Dx \ [ \ (1/2) \p_\lambda Z_p^a (D_\lambda Z_p)^a  - \p_\lambda \bar\omega^{ai} 
 \ (D_\lambda \omega_i)^a \ ].
\eeq  	
The bosonic part displays an~$O(2f)$ symmetry that acts on the index~$p$.  This symmetry group has $f(2f-1)$ real parameters whereas the $U(f)$ symmetry group previously displayed has only $f^2$ real parameters.  Thus the $O(2f)$ symmetry group is more restrictive, which reduces the number of invariant tensors. 

	The only bilinear invariants that may be formed are given by $Z_p^a Z_p^b$, which may be expressed in terms of the original fields,	
\beqa
Z_p^a Z_p^b & = & X_i^a X_i^b + Y_i^a Y_i^b
\nonumber  \\
& = & \bar\phi^{ai} \phi_i^b + \bar\phi^{bi} \phi_i^a.
\eeqa	  
The  last term corresponds to the the symmetric part of the tensor invariant (\ref{tensor}) when external sources are introduced.  The reduced variables involve the derivatives of the fields, so we replace $\bar\phi_i^a$ and $\phi_i^b$ by $\p_\lambda \bar\phi_i^a$ and $\p_\mu \phi_i^b$ in the above analysis.  Our analysis of invariants may be applied to the $O(2f)$ symmetry, with the conclusion that only the symmetric part of the tensor invariant~(\ref{tensor}) found previously,
\beq
\label{symtensor}
T_{\mu \nu}^{ab} + T_{\nu \mu}^{ba} =  \bar\psi_\mu^{aA} \psi_{\nu A}^b + \bar\psi_\nu^{bA} \psi_{\mu A}^a,
\eeq
 is a tensor invariant under the larger symmetry group.

\section{Expansion of $\Gamma^*$ about the symmetry breaking vacuum}

	There remains to express the horizon condition in terms of the reduced quantum effective action $\Gamma^*$ and the reduced variables.  By (\ref{Sigmainvariant}), we see that
\beq
{\delta \Sigma_{inv} \over \delta M}\Big|_{\Phi = 0} = {\delta \Sigma_{inv} \over \delta V}\Big|_{\Phi = 0} = {\delta \Sigma_{inv} \over \delta N}\Big|_{\Phi = 0}= {\delta \Sigma_{inv} \over \delta U}\Big|_{\Phi = 0} = 0,
\eeq
so in terms of $\Gamma^*$ the horizon condition reads, by (\ref{starsources}),	
\beqa
\label{horizonEr}
{ \delta \Gamma^* \over \delta m_{\lambda \mu}^{ab} }\Big|_{\Phi = K = L = 0} & = & { \delta \Gamma^* \over \delta v_{\lambda \mu}^{ab} }\Big|_{\Phi = K = L = 0} = 0, 
\nonumber  \\
 { \delta \Gamma^* \over \delta n_{\lambda \mu}^{ab} }\Big|_{\Phi = K = L = 0} & = & { \delta \Gamma^* \over \delta u_{\lambda \mu}^{ab} }\Big|_{\Phi = K = L = 0} = 0.
\eeqa	
This defines the symmetry-breaking vacuum.

	As in the Higgs model, to identify the Goldstone particles we expand the reduced effective action $\Gamma^*$ about the symmetry-breaking stationary point, keeping terms that are quadratic.  We are concerned with the infrared behavior, and as a first task we consider that part of $\Gamma^*$  in which there are no derivatives of the reduced variables, $A, \p c, k, L, m, n, u, v$, and which we call $\Gamma_0^*$.  
	
	We temporarily ignore the dependence of the reduced quantum effective action $\Gamma^*$ on $A$, which we may do because the global symmetries ${\cal R}_i^j$ and ${\cal F}^j$ leave $A$ invariant.  Then $\Gamma^*$ is a function of the bilinear invariant we have just found.  To expand~$\Gamma^*$ about the stationary point $\Phi = 0, Q = Q_{ph}$,  eq. (\ref{Qphys}), we need a tensor invariant that vanishes at this point, so it is a small quantity nearby.  At this stationary point, the tensor invariant we have just found has the value
\beq
(1/2)( T_{\mu \nu}^{ab} + T_{\nu \mu}^{ba})|_{\Phi = 0, Q = Q_{ph}} = - \gamma_{ph} \delta_{\mu \nu} \delta^{ab}.
\eeq
By subtracting out this term, we obtain the desired tensor invariant
\beq
s_{\mu \nu}^{ab} \equiv (1/2) (T_{\mu \nu}^{ab} + T_{\nu \mu}^{ba}) +  \gamma_{ph} \delta_{\mu \nu} \delta^{ab}.
\eeq
that satisfies
\beq
s_{\mu \nu}^{ab}|_{\Phi = 0, Q = Q_{ph}} = 0.
\eeq
It is symmetric upon interchange of both indices
\beq
s_{\mu \nu}^{ab} = s_{\nu \mu}^{ba}.
\eeq

	We expand $\Gamma_0^*$ (the part of the reduced effective action~$\Gamma^*$ that involves no derivatives of the reduced variables) in powers of this invariant tensor,
\beq
\Gamma_0^* = \int d^Dx \ [ \alpha s_{\mu \mu}^{aa} + \beta (s_{\mu \mu}^{aa})^2 + \delta s_{\mu \nu}^{ab} s_{\mu \nu}^{ab} + \epsilon s_{\mu \nu}^{ab} s_{\mu \nu}^{ba} ],
\eeq
where terms of order $s^3$ and higher are neglected.  At the stationary point (\ref{horizonEr}), we have ${\delta s \over \delta m} \neq 0$, so the first coefficient must vanish, $\alpha = 0$, and we obtain,
\beq
\label{Gammanonderiv}
\Gamma_0^* = \int d^Dx \ [ \beta (s_{\mu \mu}^{aa})^2 + \delta s_{\mu \nu}^{ab} s_{\mu \nu}^{ab} + \epsilon s_{\mu \nu}^{ab} s_{\mu \nu}^{ba} ],
\eeq
where we have neglected terms that are higher order in~$s$.

 	To find the infrared limit of the propagators, we express the reduced variables in terms of the fields.  At the stationary point $Q_{ph}$, eq.~(\ref{Qphys}), they are given by
\beqa
\label{reducedstationary}
m_\mu^{ai} & = & m_{\mu \nu}^{ab} = \gamma_{ph}^{1/2} \delta_{\mu \nu} \delta^{ab} + \p_\mu \bar\phi_\nu^{ab}
\nonumber \\
v_{\mu i}^a & = & v_{\mu \nu}^{ab} = - \gamma_{ph}^{1/2}\delta_{\mu \nu} \delta^{ab} + \p_\mu \phi_\nu^{ab}
\nonumber \\
u_\mu^{ai} & = & u_{\mu \nu}^{ab} = - \p_\mu \bar\omega_\nu^{ab}
\nonumber \\
n_{\mu i}^a & = & n_{\mu \nu}^{ab} = - \p_\mu \omega_\nu^{ab}
\nonumber \\
k_\mu^a & = & - \p_\mu \bar c^{a} ,
\eeqa
by (\ref{Qphys}) and (\ref{starsources}), where the term $\gamma^{1/2} g f^{dab} \bar\omega_\mu^{ab}$ in $k_\mu^a$ has been dropped because of the shift (\ref{cbarshift}).  This gives
\beqa
s_{\mu \nu}^{ab} & = & (1/2)(m_\mu^{ai} v_{\nu i}^b + m_\nu^{bi} v_{\mu i}^a) + ... + \gamma_{ph} \delta_{\mu \nu} \delta^{ab} 
\nonumber \\
& = & (1/2)(\gamma_{ph}^{1/2} \delta_{\mu \lambda} \delta^{ac} + \p_\mu \bar\phi_\lambda^{ac})
\nonumber \\
&&\times ( - \gamma_{ph}^{1/2} \delta_{\nu \lambda} \delta^{bc} + \p_\nu \phi_\lambda^{bc}) + ( (\mu, a) \leftrightarrow (\nu, b)) 
\nonumber  \\
&& \ \ \ \ \ \ \ \ \ \ \   + ...   + \gamma_{ph} \delta_{\mu \nu} \delta^{ab}
\nonumber \\
& = & (\gamma_{ph}^{1/2}/2)(\p_\nu \phi_\mu^{ba} - \p_\mu \bar\phi_\nu^{ab} 
+ \p_\mu \phi_\nu^{ab} - \p_\nu \bar\phi_\mu^{ba}) + ... 
\nonumber  \\
& = & (i \gamma_{ph}^{1/2}/ \sqrt 2 )( \p_\mu Y_\nu^{ab} + \p_\nu Y_\mu^{ba}) + ... \ ,
\eeqa
where the dots in the last line represent terms that are quadratic in the fields.  The leading term is linear in the fields, with coefficient $\gamma_{ph}^{1/2}$.

	We substitute this expression for $s_{\mu \nu}^{ab}$ into (\ref{Gammanonderiv}) and obtain
\beqa
\Gamma_0^* & = & { - \gamma_{ph} \over 2 } \int d^Dx  \Big(4 \beta ( \p_\mu Y_\mu^{aa})^2 + \delta ( \p_\mu Y_\nu^{ab} + \p_\nu Y_\mu^{ba})^2
\nonumber  \\
&& \ \ \ \ 
+ \epsilon ( \p_\mu Y_\nu^{ab} + \p_\nu Y_\mu^{ba})( \p_\mu Y_\nu^{ba} + \p_\nu Y_\mu^{ab}) \Big) ,
\eeqa
where $\beta, \delta, \epsilon$ are unknown  constants.  We decompose~$Y_\mu^{ab}$ into its color-symmetric and anti-symmetric parts,\footnote{For SU(2) the further decomposition of these parts into irreducible representations labeled by their dimension is given by
\beq
Y^{(ab)} = 1 \oplus 5;   \ \ \ \ \ \ Y^{[ab]} = 3
\eeq
and for SU(3) it is
\beq
Y^{(ab)} = 1 \oplus 8 \oplus 27;   \ \ \ \ \ \ Y^{[ab]} = 8 \oplus 10 \oplus \overline{10}.
\eeq}
\beqa
Y_\mu^{(ab)} \equiv (1/2) (Y_\mu^{ab} + Y_\mu^{ba})
\nonumber \\
Y_\mu^{[ab]} \equiv (1/2) (Y_\mu^{ab} - Y_\mu^{ba}),
\eeqa
which gives
\beqa
\Gamma_0^* & = & (1/2) \int d^Dx \ \Big( \beta ( \p_\mu Y_\nu^{(ab)} + \p_\nu Y_\mu^{(ab)})^2
\\ \nonumber   
&& \ \  
+ (\delta/2) ( \p_\mu Y_\nu^{[ab]} - \p_\nu Y_\mu^{[ab]})^2 + \alpha ( \p_\mu Y_\mu^{aa})^2 \Big),  
\eeqa
where $\alpha, \beta, \delta$ are (renamed) constants, and repeated indices are summed over.
To decompose into longitudinal and transverse parts, we write
\beqa
\label{effectiveaction0}
( \p_\mu Y_\nu^{(ab)} + \p_\nu Y_\mu^{(ab)})^2 = ( \p_\mu Y_\nu^{(ab)} - \p_\nu Y_\mu^{(ab)})^2 
\nonumber  \\
+ 4 \p_\mu Y_\nu^{(ab)} \p_\nu Y_\mu^{(ab)}, 
\eeqa
and obtain, after integration by parts, the final form of~$\Gamma_0^*$,
\beqa
\label{effectivepotential}
\Gamma_0^* & = & (1/2) \int d^Dx \ \Big(  (\delta/2) ( \p_\mu Y_\nu^{[ab]} - \p_\nu Y_\mu^{[ab]})^2    
   \nonumber   \\
&&  \ \ \ \
+ \beta [( \p_\mu Y_\nu^{(ab)} - \p_\nu Y_\mu^{(ab)})^2 + 4 (\p_\mu Y_\mu^{(ab)})^2]
   \nonumber   \\
&& \ \ \ \ \ \ \ \  
+  \alpha ( \p_\mu Y_\mu^{aa})^2
 \Big).
\eeqa
The only components of $\p_\mu Y_\nu^{ab}$ that are absent are the longitudinal part that is anti-symmetric in color indices,~$\p_\mu Y_\mu^{[ab]}$.

	The terms that appear in the last expression are the non-flat directions, provided that the coefficients are non-zero.  The ghost fields that do {\it not} appear are flat directions and represent Goldstone bosons and fermions. These are $c, \bar c, \omega, \bar\omega, X$,  and $\p_\mu Y_\mu^{[ab]}$.  If there are additional constraints in addition to those found here (and it should be noted that we have not implemented BRST symmetry exactly), there could be more flat directions corresponding to more Goldstone ghosts.

\section{Derivative terms in reduced effective action}

	Recall that the effective action~$\Gamma_0^*$ contains all non-derivative quadratic terms in the reduced quantum effective action~$\Gamma^*$ that are allowed by the symmetries of~$\Gamma^*$.  To evaluate the infrared limit of the propagators of fields that do not appear in~$\Gamma_0^*$, we must evaluate the quadratic terms that that involve first derivatives of the reduced variables, and that are invariant under the symmetries of~$\Gamma^*$.
	
	They are obtained by applying derivatives to the super-multiplets, out of which the bilinear invariant (\ref{tensor}) was constructed, that possesses the global $O(2f)$ symmetry, namely
\beq
\p_\kappa \bar\psi_\lambda^{aB} \ \p_\mu \psi_{\nu B}^b.
\eeq	
This is quadratic in the fields because the derivatives kill the constant term $\gamma_{ph}^{1/2}$ that appears in (\ref{reducedstationary}).  In principle we should take the symmetric part, as in (\ref{symtensor}), however only the symmetric part will contribute.  The most general color and Lorentz invariant that can be constructed  from this tensor is given by
\beq
\label{partialpsi}
\Gamma_1^* = \int d^Dx \ (\epsilon \delta_{\kappa \mu} \delta_{\lambda \nu} + \zeta \delta_{\kappa \lambda} \delta_{\mu \nu}) \ \p_\kappa \bar\psi_\lambda^{aB} \ \p_\mu \psi_{\nu B}^a ,
\eeq
where $\epsilon$ and $\zeta$ are unknown constants for which only perturbative calculations are available.  In principle there could be a third term $\delta_{\kappa \nu} \delta_{\lambda \mu}\p_\kappa \bar\psi_\lambda^{aB} \ \p_\mu \psi_{\nu B}^a$, but it may be brought into the form of the second term by partial integration.  We substitute (\ref{reducedstationary}) and obtain
\beqa
\label{partialpsia}
\Gamma_1^* & = & \int d^Dx \ (\epsilon \delta_{\kappa \mu} \delta_{\lambda \nu} + \zeta \delta_{\kappa \lambda} \delta_{\mu \nu}) 
  \nonumber  \\
&& \times ( \p_\kappa m_\lambda^{ai} \ \p_\mu v_{\nu i}^a
 - \p_\kappa u_\lambda^{ai} \ \p_\mu n_{\nu i}^a
\nonumber  \\
&& \ \ \ \ \ \ \  + \p_\kappa k_\lambda^a \ \p_\mu \p_\nu c^a)
\nonumber \\
& = & \int d^Dx \ (\epsilon \delta_{\kappa \mu} \delta_{\lambda \nu} + \zeta \delta_{\kappa \lambda} \delta_{\mu \nu}) 
\nonumber   \\
&& \times ( \p_\kappa \p_\lambda \bar\phi^{ai} \ \p_\mu \p_\nu \phi_i^a
 - \p_\kappa \p_\lambda \bar\omega^{ai} \ \p_\mu \p_\nu \omega_i^a
\nonumber  \\
&& \ \ \ \ \ \ \  - \p_\kappa \p_\lambda \bar c^a \ \p_\mu \p_\nu c^a).
\nonumber  \\
& = & \int d^Dx \ \eta  ( \p^2 \bar\phi^{ai} \ \p^2 \phi_i^a
 - \p^2 \bar\omega^{ai} \ \p^2 \omega_i^a
 \nonumber  \\
&& \ \ \ \ \ \ \ \  \ \ \ \ \    - \p^2 \bar c^a \ \p^2 c^a)
\nonumber  \\
& = & \int d^Dx \ \eta  [ (1/2)(\p^2 X_\lambda^{ab})^2 + (1/2) (\p^2 Y_\lambda^{ab})^2
 \nonumber  \\
&& \ \ \ \ \ \ \ \  \ \ \ \ \ - \p^2 \bar\omega^{ai} \ \p^2 \omega_i^a   - \p^2 \bar c^a \ \p^2 c^a].
\eeqa
where $ \eta \equiv \epsilon + \zeta $, and we have written $i = (\lambda, b)$ for the index $i$ on $X_i^a$ and $Y_i^a$.  We have obtained second derivatives of the fields, because the reduced variables $m, v, u, n$ are themselves first derivatives of the field variables.

\section{Effective action in the infrared limit}

	According to~(\ref{Sigmainvariant}), the quantum effective action is given by $\Gamma = \Sigma_{inv} + \Gamma^*$.  We keep the quadratic parts of each term,
\beq
\label{Gammaquadratic}
\Gamma_q = \Sigma_{inv,q} + \Gamma_q^*.
\eeq
The $b$-field has been integrated out, so the only remaining quadratic term in $\Sigma_{inv}$ is, by (\ref{Sigmainvariant1}),
\beqa
\label{Sigmainvariantq}
\Sigma_{inv, q} & \equiv & \int d^Dx \ \gamma_{ph}^{1/2} g A_\lambda \times (\phi - \bar\phi)_\lambda]^{aa}
\nonumber  \\
& = &\int d^Dx \ i \sqrt 2 \gamma_{ph}^{1/2} g f^{abc}A_\lambda^b \ Y_\lambda^{ca},
\eeqa

	In the reduced action $\Gamma^*$ we also expect terms of the form
\beq
\p \bar\psi^B \psi_B \ A; \ \ \ \ \ \ \  \bar\psi^B \p \psi_B \ A; \ \ \ \ \ \ \    \bar\psi^B \psi_B \  \p A.
\eeq	  
However when the physical values of the reduced variables (\ref{reducedstationary}) are substituted, they are at most of order
\beq
\p^2 \bar\phi A; \ \ \ \ \    \p^2 \phi A; \ \ \ \ \     f^{abc} \p \phi^{ab} \p A^c; \ \ \ \ \    f^{abc} \p \bar\phi^{ab} \p A^c
\eeq
and they are subleading in momentum compared to $\Sigma_{inv, q}$, and we neglect them.

	Finally, there is a term quadratic in $A$,  which we write as
\beq
\label{gamma2}
\Gamma_2^* = (1/2)(A, \kappa A),
\eeq	
where $\kappa$ has not been determined, but is restricted to either
\beq
\kappa = M^2 \ {\rm or} \ \kappa = -  b \p^2.
\eeq

	 The complete quadratic action in the neighborhood of the classical vacuum that spontaneously breaks the symmetry is given by
\beq
\Gamma_q = \Sigma_{inv, q} + \Gamma_0^* + \Gamma_1^* + \Gamma_2^*,
\eeq
where the individual terms are given in (\ref{Sigmainvariantq}), (\ref{effectivepotential}), (\ref{partialpsia}), and (\ref{gamma2}).

\section{Infrared limit of Propagators of Goldstone particles}

	Having obtained the quadratic part of the reduced quantum effective action $\Gamma^*$ in the neighborhood of the classical vacuum, we can calculate the infrared asymptotic limit of the propagators.  

	The fields $X, \omega, \bar\omega, c, \bar c$, and $\p_\mu Y_\mu^{[ab]}$, do not appear in~$\Gamma_0^*$ or~$\Sigma_{inv, q}$.  These flat directions define the Goldstone particles.  From (\ref{partialpsia}) we can immediately write the asymptotic, infrared propagators in momentum space, 
\beq
\langle c^a \bar c^b \rangle = {\delta^{ab} \over \eta (k^2)^2}; \ \ \ \  \langle \omega_i^a \bar\omega^{bj} \rangle = {\delta^{ab} \delta_i^j \over \eta (k^2)^2}.
\eeq
\beq
\label{Xprop}
\langle X_\lambda^{ab} X_\mu^{cd} \rangle = {\delta_{\lambda \mu} \delta^{ac} \delta^{bd} \over \eta (k^2)^2 }.
\eeq
They exhibit a double pole, as originally found by Gribov for the $c\bar c$ propagator in a one-loop calculation \cite{Gribov:1977wm}.  Here we find that this double pole is an exact consequence of the Goldstone mechanism which is non-perturbative.  This accords with the intuitive picture, substantiated by numerical studies \cite{GOZ:2005}, according to which the restriction to the interior of the Gribov horizon entropically favors population close to the Gribov horizon where the Faddeev-Popov operator~$M$ has its first (non-trivial) zero-eigenvalue, and thus for the fermi-ghost propagator $D_{c \bar c}(x-y) = \langle (M^{-1})_{xy} \rangle$ to be enhanced in the infrared.  The Goldstone mechanism is doing the job it should.  

	According to (\ref{effectivepotential}), the only remaining Goldstone particle is the longitudinal part of $Y_\mu^{[ab]}$.  We decompose $Y_\mu^{[ab]}$ into its transverse and longitudinal parts,
\beq
Y_\mu^{[ab]} = Y_{T, \mu}^{[ab]} + Y_{L, \mu}^{[ab]}, 
\eeq
where
\beq
Y_{L, \mu}^{[ab]} \equiv \p_\nu (\p^2)^{-1} \p_\nu Y_\nu^{[ab]}; \ \ \ \ Y_{T, \mu}^{[ab]} \equiv Y_\mu^{[ab]} - Y_{L, \mu}^{[ab]},
\eeq
The propagator of $Y_{L, \mu}^{[ab]}$ is found from the restriction of~$\Gamma_1^*$ to~$Y_{L, \mu}^{[ab]}$, namely, by (\ref{partialpsia}),
\beq
\Gamma_{1a}^* \equiv \int d^Dx \ (\eta/2) (\p^2 Y_{L, \mu}^{[ab]})^2.
\eeq 
which gives for the propagator
\beq
\label{antisymLpropagator}
\langle Y_{L, \mu}^{[ab]} Y_{L, \nu}^{[cd]} \rangle = {1  \over \eta (k^2)^2 } \ \delta^{[ab],[cd]} \ L_{\mu \nu}(k),
\eeq
where
\beq
L_{\mu \nu}(k) \equiv {k_\mu k_\nu \over k^2 }
\eeq
is the longitudinal projector, and
\beq
\delta^{[ab],[cd]} \equiv (1/2)( \delta^{ac} \delta^{bd} - \delta^{bc} \delta^{ad}),
\eeq
is the identity in the color anti-symmetric subspace.  As we shall discuss shortly, this propagator is a candidate for a carrier of a long range force.  
	
	This completes the evaluation of the propagators of the Goldstone particles.  They all have double poles $1/(k^2)^2$.

\section{Propagators of gluon non-Goldstone ghosts}
	
	The non-Goldstone particles consist of $A$, $Y_\mu^{(ab)}$, the color-symmetric part of $Y_\mu^{ab}$, and $Y_{T, \mu}^{[ab]}$, the transverse part of the color-anti-symmetric part of $Y_\mu^{ab}$.
	
	All components of the symmetric part,~$Y_\mu^{(ab)}$, appear in~$\Gamma_0^*$, the non-derivative part of~$\Gamma^*$, so the $Y_\mu^{(ab)}$ correspond to non-flat directions, and are not Goldstone particles.  The infrared limit of their propagators may be read off from $\Gamma_0^*$.  They have simple poles,
\beq
\langle Y_\mu^{(ab)} Y_\nu^{(cd)} \rangle \sim {1 \over k^2 },
\eeq
with a color and Lorentz structure that is easily obtained by inverting the color-symmetric part of (\ref{effectivepotential}), that is, the terms with coefficient $\alpha$ and $\beta$.  Because the reduced variables depend on the derivatives of the ghost fields, the Goldstone particles have double poles instead of simple poles and the non-Goldstone particles that appear only in the reduced quantum effective action have simple poles.	
	
	There remains to evaluate the propagators of $A$ and~$Y_{T, \mu}^{[ab]}$.  The color adjoint part of~$Y_{T, \mu}^{[ab]}$, defined by
\beq
Y_{T, \mu}^b \equiv (1/\sqrt N) f^{abc} Y_{T, \mu}^{[ca]},
\eeq	
mixes with $A$ in $\Sigma_{inv,q}$, and the orthogonal components of $Y_{T, \mu}^{[ab]}$ will   have simple $1/k^2$ poles.  To find these propagators, we separate out the part of the effective action that contains the gluon and ghost modes that mix,
\beqa
\label{GammaTa}
\Gamma_m & = & 
 \int d^Dx \ \Big[  i \gamma_{ph}^{1/2} g (2N)^{1/2} A_\lambda^b \ Y_{T, \lambda}^b
 \nonumber  \\
 && \ \ \ \ \   +  (\delta/2) (\p_\mu Y_{T, \nu}^a)^2
  + (1/2)(A, \kappa A)  \Big],
\eeqa
where $\delta$ comes from (\ref{effectivepotential}).  In momentum space this action corresponds to the matrix
\[
 \begin{pmatrix}
  \kappa & \quad i \gamma_{ph}^{1/2} g (2N)^{1/2} \\
& \\
  i \gamma_{ph}^{1/2} g (2N)^{1/2} & \quad  \delta k^2  \\
 \end{pmatrix}
\]
where $\kappa =  M^2$ or $\kappa = b k^2$.  The determinant of this matrix is given by
\beq
\det = \kappa \ \delta k^2 + 2N g^2 \gamma_{ph},  
\eeq
but for either value of $\kappa$, it is dominated in the infrared limit by the second term.  This gives for the transverse adjoint propagators of $A_\mu^a$ and $Y_{T, \mu}^a$  in the infrared asymptotic limit,
\beq
\label{Aprop}
\langle A_\mu^a A_\nu^b \rangle = \Big( \delta_{\mu \nu} - {k_\mu k_\nu \over k^2} \Big) \ \delta^{ab} {\delta k^2 \over 2N g^2 \gamma_{ph} }
\eeq
\beq
\langle A_\mu^a Y_\nu^b \rangle = \Big( \delta_{\mu \nu} - {k_\mu k_\nu \over k^2} \Big) \ \delta^{ab} {-i \over (2N g^2 \gamma_{ph})^{1/2} }
\eeq
\beq
\langle Y_{T, \mu}^a Y_{T, \nu}^b \rangle = \Big( \delta_{\mu \nu} - {k_\mu k_\nu \over k^2} \Big) \ \delta^{ab} {\kappa \over 2N g^2 \gamma_{ph} }.
\eeq
These three propagators are short range.  

	The gluon propagator vanishes like $k^2$ at $k = 0$.  This non-perturbative result comes from the non-renormalization of the $A$-$Y$ mixing term and, by the magic of the inverse of a $2 \times 2$ matrix, from the term~$\delta k^2$ in the quantum effective action of the $Y$-ghost. 

	We use the projector onto the color-adjoint part,
\beq
\label{adjointproject}
P^{ab, cd} \equiv (1/ N) \sum_e  f^{aeb} f^{ced}
\eeq
\beq
P^{ab,cd} P^{cd, ef} = P^{ab,ef},
\eeq
to find the propagator in the original basis,
\beqa
\label{transverseSprop}
\langle Y_{T, \mu}^{[ab]} Y_{T, \nu}^{[cd]} \rangle & = &  \Big( \delta_{\mu \nu} - {k_\mu k_\nu \over k^2} \Big) \Big( P^{ab,cd} {\kappa \over 2N g^2 \gamma_{ph} }
\\    \nonumber 
&& \ \ \ \ \ \ \ \  + ( \delta^{[ab],[cd]} - P^{ab,cd}) { 1 \over \delta k^2 } \Big),
\eeqa
\beq
\langle A_\mu^a Y_{T, \nu}^{[bc]} \rangle = \Big( \delta_{\mu \nu} - {k_\mu k_\nu \over k^2} \Big) \ {f^{abc} \over \sqrt N} {-i \over (2N g^2 \gamma_{ph})^{1/2} },
\eeq
where $\delta$ comes from (\ref{effectivepotential}).

\section{Goldstone ghosts as carriers of long-range force}  

	The $X$-bose-ghost appears only in closed loops, like the fermi-ghosts, so it cannot be exchanged between quarks.  In fact, if the $X$-field is integrated out, one obtains the factor, $(\det M)^{-f/2}$, which partially cancels the factor $(\det M)^f$ produced by the $f$ auxiliary fermi-ghost pairs $\bar\omega^i \omega_i$, where $\det M$ is the Faddeev-Popov determinant.
	
	On the other hand, the $Y$-bose-ghost mixes with the gluon field, so although it does not couple directly to a quark line, it couples to a quark line indirectly through vertex diagrams $\Gamma_\mu(p, k)$, where $p$ and $k$ are the quark and $Y$-ghost momentum respectively.  For example, a $Y$ can convert to a gluon~$A$ and another~$Y$ at the elementary vertex, $g f^{abc} Y_\nu^{bd} A_\mu^b \p_\mu Y_\nu^{cd}$.  The new $Y$ then converts to a second gluon~$A$ by a $D_{AY}$ propagator.  Both gluons are then absorbed by a quark line.  Thus the $Y$'s have an effective coupling to the color-charge of quarks.  We have found that a double pole occurs in the $Y_{L, \mu}^{[ab]}$ channel which has a piece, $Y_{L, \mu}^b = (1/\sqrt N) f^{abc} Y_{L, \mu}^{[ca]}$, in the adjoint representation.  The result is an effective quark-quark interaction given by  
\beq
\bar q \lambda^a \Gamma_\mu(p, k) q \ {k_\mu k_\sigma \over k^2} \ {1 \over \eta (k^2)^2} \ \bar q' \lambda^a \Gamma_\sigma(p', k) q'.
\eeq
	
	At first sight it appears that the double pole corresponds to a linearly rising potential between quarks.  However in Landau gauge, an external ghost $k$ momentum factors out of every vertex diagram with an external ghost line,\footnote{This may be seen from (\ref{starsources}) where only the derivatives $\p \phi$ and $\p \bar\phi$ appear in the reduced variables $m$ and $v$.} so the $Y$-ghost-quark vertex is of the form
\beq
\label{factextghost}
\Gamma_\mu(p, k) = H_{\mu \nu}(p, k) k_\nu,
\eeq
where
\beq	
H_{0, \mu \nu}(p) \equiv H_{\mu \nu}(p, 0),
\eeq
is finite at $k = 0$.  Thus the quark-quark interaction due to exchange of a single~$Y$ quantum is given, at small $k$, by
\beq
\bar q \lambda^a H_{0, \mu \nu}(p) q \ {k_\mu k_\nu \over k^2} \ {1 \over k^2} \ { k_\sigma k_\tau \over k^2 } \ \bar q' \lambda^a H_{0, \sigma \tau}(p') q',
\eeq
The on-shell form factor, with $p^2 = m^2$ and ${p'}^2 = {m'}^2$, and with Dirac spinors satisfying
\beq
\gamma \cdot p \ q(p) = m \ q(p); \ \ \ \    \bar q(p+k) \ \gamma \cdot (p+k) = m \ \bar q(p+k)
\eeq 
is given at small $k$, by
\beq	
\bar q \lambda^a H_{0, \mu \nu}(p) q \ k_\mu k_\nu = \bar q \lambda^a q \ [c_1 (p \cdot k)^2 + c_2 k^2].
\eeq
Thus the exchange of a $Y$-type Goldstone boson between quarks results in the effective quark-quark interaction
\beq
\label{quarkexchange}
\bar q \lambda^a q \ { [c_1 (p \cdot k)^2 + c_2 k^2]    [c'_1 (p' \cdot k)^2 + c'_2 k^2] \over (k^2)^3 } \ \bar q' \lambda^a q'.
\eeq

	This interaction is of order $1/k^2$ and does not represent a linearly rising potential between quarks.  Nevertheless it does correspond to one-particle exchange between quarks of a massless quantum in the color-adjoint representation.  A further non-perturbative analysis, which we do not attempt here, would be required to determine if it can be the origin of a confining force.  Moreover, as we have noted,  there could be additional flat directions corresponding to additional Goldstone bosons in the $Y_T$ channel, which would modify this effective quark-quark interaction.  We shall return to this possibility in the concluding section.  A one-loop calculation of the effective potential between quarks using the present action is reported in~\cite{Gracey:0909}.

\section{Symmetries of the Lagrangian density}

	This section offers a possible interpretation of the spontaneous symmetry breaking we have found, but it does not modify the calculation reported here.

	Recall that the symmetry-breaking we have found occurs spontaneously in the quantum effective action~$\Gamma(\Phi, Q)$ when the sources~$Q$ are transformed appropriately.  Indeed, we have seen in sect.~\ref{symGammasymS0} that $\Gamma(\Phi, Q)$ obeys Ward identities that express the global symmetries~${\cal R}_i^j$,  and~${\cal F}^j$, and the  BRST-symmetry~$s$ of $S_0$.  These symmetries are explicitly, though softly, broken by the dimension 2 term ${\cal L}_\gamma$ in the local action $S$.  It is natural to ask, ``Could these symmetries of the quantum effective action $\Gamma$ also be symmetries of the local action~$S$?"
	
	To make the question precise, we quantize in a periodic Euclidean box.  In this case, the answer to the question is ``No, $s$ and ${\cal R}_i^j$,  and ${\cal F}^j$ are not symmetries of the action~$S$.  But they are symmetries of the Lagrangian density ${\cal L}$ that hold locally, within each coordinate patch in which the Cartesian coordinates $x_\mu$ are well defined."  (The coordinates $x_\mu$ are not well defined globally on a periodic box because they are not periodic.)
	
	Within such a coordinate patch, the Lagrangian density~${\cal L}  = {\cal L}_0 + {\cal L}_\gamma$ at finite Gribov mass~$\gamma$ may be obtained from~${\cal L}_0$ at $\gamma = 0$ by the change of variable~\cite{Schaden:1994},
\beq
\label{LzerotoL}
{\cal L}(\phi_\mu, \bar\phi_\nu, b, \bar c) = {\cal L}_0(\varphi_\mu, \bar\varphi_\nu, b^\star, \bar c^\star),  
\eeq
where
\beqa
\label{starvariables}
\varphi_\mu^{ab}  & \equiv & \phi_\mu^{ab} - \gamma^{1/2} x_\mu \delta^{ab}
 \nonumber  \\
\bar\varphi_\mu^{ab} & \equiv &   \bar\phi_\mu^{ab} + \gamma^{1/2} x_\mu \delta^{ab}
\nonumber  \\
b^{\star d}  &  \equiv & b^d + i \gamma^{1/2} g f^{adb} x_\mu \bar\phi_\mu^{ba} 
\nonumber  \\
\bar c^{\star d} & \equiv &  \bar c^d + \gamma^{1/2} g f^{adb} x_\mu \bar\omega_\mu^{ba}, 
\eeqa
and all other field variables are unchanged.  By this change of variable, each symmetry $\hat X$ of~${\cal L}_0$ is translated into a symmetry of~${\cal L}$.  A thorough analysis of the symmetries of ${\cal L}_0$ is presented in~\cite{Schaden:1994}.  

	The change of variable contains ~$x_\mu$ explicitly, so it is not translation invariant.  Nevertheless both local Lagrangian densities ${\cal L}_0$ and ${\cal L}$ are translation-invariant.  This happens because both Ldagrangian densities are invariant under shift of $\phi$ and $\bar\phi$ by constants,
\beq
\phi_\mu^{ab} \to \phi_\mu^{ab} + a_\mu \delta^{ab}; \ \ \ \ \ \ \ \ \    \bar\phi_\mu^{ab} \to \bar\phi_\mu^{ab} + \bar a_\mu \delta^{ab},
\eeq  
with a compensating shift of $b$ and $\bar c$.  We have not considered this invariance explicitly because it is implicit in the solution of the equations of motion of the ghost fields given in Appendix~A that is used in~(\ref{starsources}). 
 
	We exhibit the BRST operator  that is a symmetry of~$\cal L$.  An alternative non-local BRST operator that is a symmetry of $S$ may be found in \cite{Sorella:0905}.  For this purpose we introduce the operator $\hat s$ that acts on the Faddeev-Popov fields in the usual way
\beq
\hat s A \equiv Dc;\ \ \ \ \   \hat s c = - (g/2) c \times c; \ \ \ \ \  \hat s \bar c^\star = i b^\star; \ \ \ \ \   \hat s b^\star = 0, 
\eeq
and that acts on the (untransformed) auxiliary ghosts according to
\beqa
\hat s \varphi_\mu^{ab} & = & \omega_\mu^{ab}; \ \ \ \ \ \ \ \ \ \ \ \ \ \ \ \ \ \ \   \hat s \omega_\mu^{ab} = 0
\nonumber  \\
\hat s \bar\omega_\mu^{ab} & = & \bar\varphi_\mu^{ab}; \ \ \ \ \ \ \ \ \ \ \ \ \ \ \ \ \ \ \   \hat s \bar\varphi_\mu^{ab} = 0.
\eeqa
It is a symmetry of ${\cal L}_0(\varphi_\mu, \bar\varphi_\nu, b^\star
, \bar c^\star)$,
\beq
\hat s {\cal L}_0(\varphi_\mu, \bar\varphi_\nu, b^\star
, \bar c^\star) = 0.
\eeq
Under the change of variable (\ref{starvariables}), $\hat s$ acts on the transformed fields according to
\beqa
\hat s \phi_\mu^{ab} & = & \omega_\mu^{ab}; \ \ \ \ \ \ \ \ \ \ \ \ \ \ \ \ \ \ \ \ \ \ \  \hat s \omega_\mu^{ab} = 0
\nonumber  \\
\hat s \bar\omega_\mu^{ab} & = & \bar\phi_\mu^{ab} + \gamma^{1/2} x_\mu \delta^{ab} ; \ \ \ \ \ \ \  \hat s \bar\phi_\mu^{ab} = 0,
\eeqa
and the action on the other fields is unchanged.  Here the fields are evaluated within a coordinate patch at the point $x$, so $\bar\omega = \bar\omega(x)$, etc.\   Since this is merely a change of variable, $\hat s$ is a symmetry of ${\cal L}(\phi_\mu, \bar\phi_\nu, b, \bar c)$,
\beq
\hat s {\cal L}(\phi_\mu, \bar\phi_\nu, b, \bar c) = 0.
\eeq

	The symmetry $\hat s$ is spontaneously broken  because the vacuum state satisfies
\beq
\langle \hat s \bar\omega_\mu^{ab} \rangle = \langle \ (\bar\phi_\mu^{ab} + \gamma^{1/2} x_\mu \delta^{ab}) \  \rangle = \gamma^{1/2} x_\mu \delta^{ab} \neq 0,
\eeq
so the expecation-value of an $\hat s$-exact quantitiy is non-zero. 

	Because of the Ward identities satisfied by the auxiliary ghosts, the reduced quantum effective action 
\beq
\Gamma^*(m, v, ... ) = \Gamma^*(M + \p \bar\phi, V + \p \phi, ...)
\eeq	
depends only on the combinations $m = M + \p \bar\phi$ and $v = V + \p \phi$.  The symmetry-breaking vacuum is given by $m = - v = \gamma^{1/2}$, and we may equivalently attribute~$\gamma^{1/2}$ either to the sources $M = - V = \gamma^{1/2}$, as we have done previously, or to the fields $\varphi_\mu^{ab} = - \bar\varphi_\mu^{ab} = \gamma^{1/2} x_\mu \delta^{ab}$.  If it is attributed to the sources, then~${\cal L}$ breaks the symmetry explicitly but softly; if it is attributed to the fields then~${\cal L}$ breaks the symmetry spontaneously. 

	Although these two points of view are strictly equivalent for the purpose of calculating the propagators, it may be helpful, when considering the problem of unitarity and positivity of the present approach, to consider the BRST symmetry as being spontaneously broken. 
	
	Finally we note that a sufficient condition for the existence of a conserved BRST Noether current is that the symmetry operator~$\hat s$ be well defined within a coordinate patch.  To exhibit this current, consider the infinitesimal variation
\beq
\delta \Phi_i = \epsilon(x) \hat s \Phi_i,
\eeq	
where $\epsilon(x)$ vanishes outside a coordinate patch on which the coordinate~$x_\mu$ is well defined, but is an otherwise arbitrary function of~$x$.  Since~$\hat s$ is a symmetry of the Lagrangian density, we have, by Noether's theorem,
\beq
\delta {\cal L} = j_\mu \p_\mu \epsilon,
\eeq
where
\beq
j_\mu = {\p {\cal L} \over \p \p_\mu \Phi_i } \hat s \Phi_i 
\eeq
is the conserved BRST current.

\section{Discussion}

	In the present work we have shown that the fermi ghosts and some bose ghosts are the Goldstone particles of a symmetry that is spontaneously broken at the level of the quantum effective action $\Gamma$.  Their propagators possess a double pole.  
\beq
D_{c \bar c} = D_{\omega \bar\omega} = D_{X X} \sim D_{Y_L^{[ab]} Y_L^{[cd]}} \sim {1 \over (k^2)^2}.
\eeq	
The double pole in the fermi ghost propagator agrees with the Kugo-Ojima confinement criterion \cite{Kugo-Ojima}.  The relation between the present approach and the Kugo-Ojima approach has been clarified recently~\cite{Dudal:1001}.
	
	We have not fully exploited the Slavnov-Taylor identity because of its non-linearity, and in principle there could be other symmetries, not considered here, that further constrain $\Gamma$.  Thus there may be additional flat directions besides the ones whose existence we have established, and corresponding additional Goldstone particles.  These could only§ be in the $Y$ propagator, because the propagators of all other ghosts have double poles.  In this respect a comparison with the recent perturbative calculations of Gracey~\cite{Gracey:0909} is illuminating.  Starting from the present action, Gracey has calculated the ghost and gluon propagators to one-loop order and imposed the horizon condition to this order.  Although our calculation of the infrared limit of propagators is non-perturbative, the symmetries we have found hold order by order in perturbation theory, and the Goldstone particles we have found should be seen in each order of perturbation theory when the horizon condition is imposed.  Indeed, the double poles we have found the the $c$-$\bar c, \omega$-$\bar\omega, X,$ and $Y_L^{[ab]}$ propagators  also appear in Gracey's calculation, as does the suppression of the gluon propagator.\footnote{Gracey's result for the longitudinal part of the $Y$-$Y$ propagator (the $\xi$-$\xi$ propagator in his notation) is not reported in \cite{Gracey:0909}.  I am grateful to him for communicating this result to me privately.}  However he finds additional double poles in the transverse part of  the $Y$-$Y$ propagator that we have not found.  This suggests that further exploitation of the Slavnov-Taylor identity would reveal additional Goldstone bosons corresponding to the additional double poles found in one-loop by Gracey.  If present, they would modify the quark-quark effective interaction given in~(\ref{quarkexchange}).
	
	We have found as an exact result, eq.~(\ref{Aprop}), that the gluon propagator $D(k)$ vanishes like $k^2$ at $k = 0$, 
\beq
D_{AA}(k) \sim k^2,
\eeq	
as originally found by Gribov~\cite{Gribov:1977wm}.  This nicely explains the absence of gluons from the physical spectrum.  Indeed, the equation $D(0) = 0$ is not compatible with the Lehmann representation, 
\beq
\label{lehmann}
D(k) = \int_0^\infty dM^2 \rho(M^2)/(k^2 + M^2),
\eeq
and a positive spectral function $\rho(M^2) \geq 0$.  However the short range of the gluon propagator only deepens the mystery, in the Gribov approach in Landau gauge, of the origin of the long-range confining force between quarks.  Indeed, in his original paper, Gribov turned from the Landau to the Coulomb gauge to address this problem \cite{Gribov:1977wm}.  The exchange of a massless quantum between quarks that is assured by the Goldstone mechanism exhibited here may offer a resolution of this dilemma.  This possible confinement mechanism is also proposed in~\cite{Gracey:0909}.
	
	From the results $D_{AA}(k) \sim k^2$ and $D_{c \bar c}(k) \sim 1/ (k^2)^2$, where~$D_{c \bar c}$ is the Faddeev-Popov ghost propagator, we obtain for the renormalization-group invariant running effective coupling constant \cite{Alkofer:2000},
\beq
\alpha_s^{\rm eff}(k) \equiv (g^2 / 4 \pi) (k^2)^3 D_{AA}(k) D_{c \bar c}^2(k),
\eeq	
the finite infrared limit
\beq
\alpha_s^{\rm eff}(0) = O(1),
\eeq
in agreement with the one-loop result~\cite{Gracey:0909}.
	
	It was recently proposed \cite{Zwanziger:0904} that there is a triple pole in the Y-propagator which, with the factorization of ghost momentum from quark-ghost vertex, eq.~(\ref{factextghost}), gives an effective one-particle exchange between quarks of the form~$1/(k^2)^2$ that could be the carrier of a linearly rising potential between quarks.  Although a triple pole is not indicated by the present calculation, it remains a possibility if the coefficient of the $(k^2)^2$ term in the effective action of the relevant $Y$ ghost were to vanish.  In any case, we also find here that the longest range force between quarks arises from exchange of a Y-quantum.
	 
	The infrared exponents (infrared power laws of the propagators) that we have obtained are integer, as they are in finite-order of perturbation theory \cite{Gracey:0909}.  In contrast, a recent solution of the Dyson-Schwinger equation derived from the local action used here \cite{Huber:0910},  gives non-integer infrared exponents.  This difference may have its origin in truncation error in the Dyson-Schwinger calculation, or possibly to different choice of Gribov copy (different gauge) inside the Gribov horizon, or in the fact that the behavior obtained here is delicate because it holds only at $\gamma = \gamma_{ph}$, and could be missed in a Dyson-Schwinger calculation.  It would be of interest to look for solutions to the Dyson-Schwinger equations that have the infrared limit that has been found here.
			   	
	We now turn to a comparison with lattice data, which has been reviewed recently in~\cite{Cucchieri:2010.03}.  Recall that we have obtained a gluon propagator that vanishes like~$k^2$, independent of dimension.  In 2 Euclidean dimensions it is found numerically that $D(0) = 0$, in accordance with this result, but numerically it appears that $D(k)$ vanishes like $k^p$, with $p < 2$~\cite{Maas:2007}.  In contrast, on large lattices in~3 and~4 Euclidean dimensions, it appears that the gluon propagator is finite, $D(0) > 0$, at $k = 0$ \cite{Cucchieri:2008.12, Cucchieri:2008.04, Ilgenfritz:2009, Oliveira:2008, vonSmekal2007, Cucchieri:2007b, Maas:2008}.  In~3 dimensions there is a clear turnover of $D(k)$ which has a maximum at finite $k$, and approaches its infrared limit,~$D(0)$, from above, ${\p D(k) \over \p k} > 0$ at low~$k$.  So in 3 dimensions $D(0)$ though finite, is suppressed at $k = 0$~\cite{Maas:2008, Cucchieri:2010.03}.  Note that from the Lehmann representation (\ref{lehmann}), the first deriveative of $D(k^2)$,
\beq	
{\p D(k^2) \over \p k^2} = - \int_0^\infty dM^2 { \rho(M^2) \over (k^2 + M^2)^2},
\eeq
is negative if the spectral function is positive, $\rho(M^2) \geq 0$.  Thus the observed turnover of $D(k)$ in 2 and 3 Euclidean dimensions, with ${\p D(k) \over \p k} > 0$ at low $k$, implies that the gluon field produces unphysical excitations.  In $d = 4$ dimensions there appears to be a shoulder in $D(k)$, if not a turn-over, and this would also contradict the Lehmann representation with positive spectral function, because every derivative $\p^n D(k^2) \over (\p k^2)^n$ is monotonic.  

	The only explanation at hand for the observed turnover of the gluon propagator in 2 and 3 dimensions is the suppression of infrared modes due to the proximity of the Gribov horizon in infrared directions, and which is otherwise counter-intuitive.  It is puzzling that the main qualitative feature of the Gribov scenario is confirmed by latttice studies, namely suppression of infrared gluon modes, but numerically there is disagreement with lattice data at $k = 0$.  The situation with the ghost propagator is similar.  We have found that the fermi-ghost propagator has a double pole $1/(k^2)^2$, corresponding to a dressing function $k^2 G(k)$ that diverges at $k = 0$.  However lattice data show a ghost dressing function that does increase as~$k$ descreases, but which appears to level off at the lowest~$k$ available~\cite{Cucchieri:2010.03}.
	
	A possible way out has been proposed by Maas~\cite{Maas:2008}.  He has studied numerically the properties of different Gribov copies inside the Gribov horizon, and found that, depending on the choice of weight given to different copies, one may impose any one of a continuum of values for~$D(0)$, the gluon propagator at $k = 0$.  This effectively makes~$D(0)$ into a gauge parameter, within the class of Landau gauges inside the Gribov horizon.  Thus it is possible that the results obtained here correspond to a particular Landau gauge within the Gribov horizon.  An alternative approach which accords with the lattice data is to modify the local action used here to account for condensation of dimension-2 operators~\cite{Dudal:2008sp}.
	
	 Other unresolved questions are the identification of the physical states and observables.  It has been found that renormalization of the operator $F^2$ requires both BRST exact and BRST non-invariant quantities to construct a quantum operator invariant under renormalization-group equations \cite{Dudal:0908}.  A possible construction of physical observables is developed in \cite{Baulieu:2009ha}.  We remain far from a satisfactory understanding of the phases of QCD.

\bigskip
	  
{\bf Acknowledgements}\\
The author recalls with pleasure stimulating conversations with Reinhard Alkofer, Laurent Baulieu, David Dudal, John Gracey, Klaus Lichtenegger, Jan Pawlowski, Valentin Reys, Alexander Rutenburg, Martin Schaden, Silvio Sorella, Nele Vandersickel, and Lorenz von Smekal.

\appendix

\section{Solution of ghost equations of motion }

	The extended action with sources for composite operators is given in (\ref{extendaction}).  In this Appendix we shall use the 4 sources $M, N, U, V$ to convert the equations of motion of the 4 auxiliary ghosts into 4 Ward identities that are stable under renormalization.  This was done in \cite{Zwanziger:1993}, but we report it here for completeness.  A new result reported in this Appendix is the solution of the integrated equation of motion of the Faddeev-Popov ghost~$c$.

	But first let us recall the Ward identities associated with the Faddeev-Popov fields $b$ and $\bar c$,
\beq
{\delta \Gamma \over \delta b} = - i \p_\lambda A_\lambda; \ \ \ \ \ \ \ \   {\delta \Gamma \over \delta \bar c} = \p_\lambda {\delta \Gamma \over \delta K_\lambda}, 
\eeq	
which have the solution
\beq
\Gamma = \Gamma_1(K_\lambda - \p_\lambda \bar c, ...) + i(\p_\lambda b, A_\lambda).
\eeq
This gives the complete dependence of $\Gamma$ on $b$ and $\bar c$.  We now derive similar equations for the 4 auxiliary ghosts.

	We have
\beqa
\label{omegaEOMa}
{\delta \Sigma \over \delta \bar\omega^{ai} } = & \p_\lambda [ s (D_\lambda \phi_i)^a ] &  - (D_\lambda N_{\lambda i})^a
\nonumber  \\
&&+ g f^{abc} (D_\lambda c)^b V_{\lambda i}^c.
\eeqa
We use the identity
\beq
{ \delta \Sigma \over \delta U_\lambda^{ai} } = 
s (D_\lambda \phi_i)^a - N_{\lambda i}^a
\eeq
to write this as
\beqa
\label{omegaEOMa}
{\delta \Sigma \over \delta \bar\omega^{ai} } & = & \p_\lambda {\delta \Sigma \over \delta U_\lambda^{ai} } - (g A_\lambda \times N_{\lambda i})^a
\nonumber  \\
&& \ \ \ \ \ \ \ \ \ \ \ \ \ \ \ \ \ \ \ \ \ \ \  
+ g f^{abc} {\delta \Sigma \over \delta K_\lambda^b} V_{\lambda i}^c.
\eeqa
This equation is at most linear in the functional derivatives and in the field $A_\lambda$, and consequently one may show, by the method that was used to establish the Slavnov-Taylor identity, that the same equation is satisfied by the quantum effective action~$\Gamma$, 
\beqa
\label{omegaEOMb}
{\delta \Gamma \over \delta \bar\omega^{ai} } & = & \p_\lambda {\delta \Gamma \over \delta U_\lambda^{ai} }  - (g A_\lambda \times N_{\lambda i})^a
\nonumber  \\
&& \ \ \ \ \ \ \ \ \ \ \ \ \ \ \ \ \ \ \ \ \ \ \  
+ g f^{abc} {\delta \Gamma \over \delta K_\lambda^b} V_{\lambda i}^c.
\eeqa
This is the first Ward identity.  It has the solution
\beqa
\Gamma = \Gamma_2 + (i \p_\lambda b, A_\lambda) + (g A_\lambda \times \bar\omega^i, N_{\lambda i}),
\eeqa
where
\beq
\label{solomegabar}
\Gamma_2 \equiv \Gamma_2(U_\lambda^i - \p_\lambda \bar\omega^i, K_{1,\lambda}),
\eeq
\beq
\label{kayone}
K_{1, \lambda}^a \equiv K_\lambda^a - \p_\lambda \bar c^a - g  (\bar\omega^i \times V_{\lambda i})^a ,
\eeq
and $(\bar\omega^i \times V_{\lambda i})^a \equiv f^{abc} \bar\omega^{bi} V_{\lambda i}^c$.  This gives the complete dependence on $\bar\omega$.

	Likewise we have
\beq
\label{phiEOM}
{\delta \Sigma \over \delta \bar\phi^{ai} } = - \p_\lambda (D_\lambda \phi_i)^a 
  - (D_\lambda V_{\lambda i})^a,
\eeq	
We use
\beq
{\delta \Sigma \over \delta M_\lambda^i } = D_\lambda \phi_i + V_{\lambda i}
\eeq
to write this as
\beq
\label{phiEOMa}
{\delta \Sigma \over \delta \bar\phi^{ai} } = - \p_\lambda {\delta \Sigma \over \delta M_\lambda^{ai} } 
  - (g A_\lambda \times V_{\lambda i})^a.
\eeq
Again this is at most linear in the derivatives and the field $A_\mu$, so $\Gamma$ satisfies the same equation
\beq
\label{phiEOMb}
{\delta \Gamma \over \delta \bar\phi^{ai} } = - \p_\lambda {\delta \Gamma \over \delta M_\lambda^{ai} }  
  - (g A_\lambda \times V_{\lambda i})^a,
\eeq
which is the second Ward identity.  It has the solution
\beqa
\Gamma & = & \Gamma_3 + (i \p_\mu b, A_\mu)
\nonumber  \\
&& + (g A_\lambda \times \bar\phi^i, V_{\lambda i}) + (g A_\lambda \times \bar\omega^i, N_{\lambda i}),
\eeqa
where
\beq
\Gamma_3  \equiv \Gamma_3(M_\lambda^i + \p_\lambda \bar\phi, U_{\lambda \mu} - \p_\lambda \bar\omega^i, K_{1, \lambda}) ,
\eeq
and we have made use of our previous result.  This gives the complete dependence on $\bar\omega$ and $\bar\phi$.

	To derive the third Ward identity, we start from
\beqa
{\delta \Sigma \over \delta \omega_i^a} & = & - (D_\lambda \p_\lambda \bar\omega^i)^a + (D_\lambda U_\lambda)^{ai}
\\   \nonumber
& = & - \p_\lambda (\D_\lambda \bar\omega^i)^a  + g (\p_\lambda A_\lambda \times \bar\omega^i)^a + (D_\lambda U_\lambda)^{ai}.
\eeqa
We use
\beq
{\delta \Sigma \over \delta N_{\lambda i} } = - D_\lambda \bar\omega^i + U_\lambda^i
\eeq
to write this as
\beq
{\delta \Sigma \over \delta \omega_i^a}
= \p_\lambda {\delta \Sigma \over \delta N_{\lambda i}^a } +  ig \Big({\delta \Sigma \over \delta b} \times \bar\omega^i\Big)^a + (g A_\lambda \times U_\lambda^i)^a.
\eeq
One can show, using the equation of motion of $b$ that the quantum effective action satisfies the same equation, 
\beq
{\delta \Gamma \over \delta \omega_i^a} 
= \p_\lambda {\delta \Gamma \over \delta N_{\lambda i}^a }   +  ig \Big({\delta \Gamma \over \delta b} \times \bar\omega^i\Big)^a
 + (g A_\lambda \times U_\lambda^i)^a,
\eeq
which is the third Ward identity.  It may also be written
\beq
{\delta \Gamma \over \delta \omega_i^a} 
= \p_\lambda {\delta \Gamma \over \delta N_{\lambda i}^a }   +   ( g \p_\lambda A_\lambda \times \bar\omega^i)^a
 + (g A_\lambda \times U_\lambda^i)^a,
\eeq
which has the solution
\beq
\Gamma = \Gamma_4(N_{\lambda i} - \p_\lambda \omega_i) + (\bar\omega^i, g \p_\lambda A_\lambda \times \omega_i) + (U_\lambda^i, g A_\lambda \times \omega_i).
\eeq
This gives the complete dependence on $\omega$.  We wish to write this solution for $\omega$ in a way that is compatible with our previous solution for $\bar\omega$.  To this end we write
\beqa
\Gamma_4(N_{\lambda i} - \p_\lambda \omega_i) & = & \Gamma_5(N_{\lambda i} - \p_\lambda \omega_i) 
\nonumber   \\
&& \ \ \ \ \ \  + (g A_\lambda \times \bar\omega^i, N_{\lambda i} - \p_\lambda \omega_i)
\nonumber  \\ \nonumber
& = & \Gamma_5(N_{\lambda i} - \p_\lambda \omega_i) + (g A_\lambda \times \bar\omega^i, N_{\lambda i})
\\  \nonumber  
&& - (\p_\lambda \bar\omega^i, g A_\lambda \times \omega_i)
\nonumber  \\
&& \ \ \ \ \ \ \ \ \ 
 - (\bar\omega^i, g \p_\lambda A_\lambda \times \omega_i).
\eeqa
The terms in $\p_\lambda A_\lambda$ cancel and we obtain
\beqa
\Gamma & = & \Gamma_5(N_{\lambda i} - \p_\lambda \omega_i) + (U_\lambda^i - \p_\lambda \bar\omega^i, g A_\lambda \times \omega_i)
\nonumber  \\
&& \ \ \ \ \ \ \  +  (g A_\lambda \times \bar\omega^i, N_{\lambda i}).
\eeqa
This expression is compatible with the previous solution for $\bar\omega$ (and $\bar\phi$), and we obtain
\beqa
\label{omegasol}
\Gamma & = & \Gamma_6  + (i \p_\mu b, A_\mu) + (U_\lambda^i - \p_\lambda \bar\omega^i, g A_\lambda \times \omega_i) 
\nonumber  \\
&& \ \ \    + (g A_\lambda \times \bar\phi^i, V_{\lambda i}) + (g A_\lambda \times \bar\omega^i, N_{\lambda i}),
\eeqa
where
\beq
\Gamma_6 \equiv \Gamma_6({M'_\lambda}^i,  N'_{\lambda i},  {U'_\lambda}^i, K_{1, \lambda}),
\eeq
and the primed variables are defined below in (\ref{primesources}).  This gives the complete dependence on $\bar\omega, \bar\phi$, and $\omega$.

	To derive the fourth Ward identity we start from
\beqa
\label{phibarEOMc}
{\delta \Sigma \over \delta \phi_i^a } & = & - (D_\lambda \p_\lambda \bar\phi^i)^a + [(U_\lambda^i - \p_\lambda \bar\omega^i) \times g D_\lambda c ]^a 
\nonumber \\
&&  \ \ \ \ \ \ \ \ \ \  - (D_\lambda M_\lambda^i)^a
\nonumber  \\
& = & - \p_\lambda ( s D_\lambda \bar\omega^i)^a + g ( \p_\lambda A_\lambda \times  \bar\phi^i)^a 
\nonumber \\
&& \ \ \ \ \ \ + g (\p_\lambda D_\lambda c \times \bar\omega^i)^a - (D_\lambda M_\lambda^i)^a 
\nonumber  \\
&& \ \ \ \ \ \ \  + (U_\lambda^i \times g D_\lambda c )^a.
\eeqa
We use the identity,
\beq
{\delta \Sigma \over \delta V_{\lambda i} } = s D_\lambda \bar\omega^i + M_\lambda^i,
\eeq
to write this as
\beqa
\label{phibarEOMd}
{\delta \Sigma \over \delta \phi_i^a } & = & - \p_\lambda { \delta \Sigma \over \delta V_{\lambda i}^a }
+ ig \Big( {\delta \Sigma \over \delta b } \times \bar\phi^i \Big)^a  - (g A_\lambda \times M_\lambda^i)^a
\nonumber  \\
&& + g \Big(\p_\lambda {\delta \Sigma \over \delta K_\lambda } \times \bar\omega^i \Big)^a
   + \Big( g U_\lambda^i \times {\delta \Sigma \over \delta K_\lambda } \Big)^a.
\eeqa
The quantum effective action satisfies the same equation, which yields the fourth Ward identity,
\beqa
\label{phibarEOMe}
{\delta \Gamma \over \delta \phi_i^a } & = & - \p_\lambda { \delta \Gamma \over \delta V_{\lambda i}^a }
+ ig \Big( {\delta \Gamma \over \delta b } \times \bar\phi^i \Big)^a - (g A_\lambda \times M_\lambda^i)^a
\nonumber  \\
&& + g \Big( \p_\lambda {\delta \Gamma \over \delta K_\lambda } \times \bar\omega^i \Big)^a
   + \Big( g U_\lambda^i \times {\delta \Gamma \over \delta K_\lambda } \Big)^a  .
   \eeqa
This has the solution
\beq
\label{phisol1}
\Gamma = \Gamma_7 - (\bar\phi^i, g \p_\lambda A_\lambda \times \phi_i)
 +(M_\lambda^i, g A_\lambda \times \phi_i),
\eeq
where
\beq
\Gamma_7 = \Gamma_7(V_{\lambda i} + \p_\lambda \phi_i, K_{2, \lambda}),
\eeq
\beqa
K_{2, \lambda} & = &  K_\lambda + \p_\lambda(g \bar\omega^i \times \phi_i) - g U_\lambda^i  \times \phi_i
  \\
& = & K_\lambda + g \bar\omega^i \times \p_\lambda\phi_i - g (U_\lambda^i - \p_\lambda\bar\omega^i)  \times \phi_i
\nonumber  
\eeqa
and $(\bar\omega^i \times \phi_i)^a = f^{abc}\bar\omega^{bi} \times \phi_i^c$ etc.  This gives the complete $\phi$ dependence.  To make the solution for $\phi$ compatible with the solution for $\bar\phi$ and $\bar\omega$, we make use of the fact that the dependence on $V_{\lambda i} + \p_\lambda \phi_i$ is completely arbitrary, and  moreover, in the solution for $\phi$ we may freely choose the dependence on the variables $\bar\phi, \bar\omega, \omega$ to be consistent with our previous solution.  We write
\beqa
\label{gamma7to8}
\Gamma_7(V_{\lambda i} + \p_\lambda \phi_i, K_2)  \equiv ( g A_\lambda \times \bar\phi^i, V_{\lambda i} + \p_\lambda \phi_i)
\nonumber  \\
 + \Gamma_8(V_{\lambda i} + \p_\lambda \phi_i, K')
 \nonumber  \\
 = ( g A_\lambda \times \bar\phi^i, V_{\lambda i}) + (\p_\lambda \bar\phi^i, g A_\lambda \times \phi_i) 
\nonumber  \\
 - ( g \p \cdot A \times \bar\phi^i, \phi_i) + \Gamma_8(V_{\lambda i} + \p_\lambda \phi_i, K'),
\eeqa
where
\beq
K'_\lambda \equiv K_{2, \lambda} - \p_\lambda \bar c - g \bar\omega^i \times (V_{\lambda i} + \p_\lambda \phi_i)
\eeq
is given below in (\ref{primesources}).  The dependence of $K'$ on $\bar\omega$ is now consistent with (\ref{solomegabar}) and (\ref{kayone}).  We substitute (\ref{gamma7to8}) into (\ref{phisol1}).  The terms in $\p \cdot A$ cancel,  and we obtain
\beqa
\Gamma & = & (\p_\lambda \bar\phi^i + M_\lambda^i, g A_\lambda \times \phi_i) 
\nonumber  \\
&& \ \ \ \   + ( g A_\lambda \times \bar\phi^i, V_{\lambda i})  + \Gamma_8(V', K'),
\eeqa
where the primed variables are given in (\ref{primesources}).  This expression gives the complete dependence of $\Gamma$ on $\phi$ and moreover it is compatible with our previous solution (\ref{omegasol}) for $\bar\omega, \bar\phi$, and $\omega$.  We combine the two expressions and obtain 
\beq
\label{primesolution}
\Gamma = \Sigma_p + \Gamma'(A, c, K', L, M', N', U', V'),
\eeq
where
\beqa
\label{Sigmaprime}
\Sigma_p \equiv (i \p_\lambda b, A_\lambda) 
 + (M_\lambda^i + \p_\lambda \bar\phi^i, g A_\lambda \times \phi_i)
 \nonumber \\ 
+ (g A_\lambda \times \bar\phi^i, V_{\lambda i}) + ( g A_\lambda \times \bar\omega^i, N_{\lambda i}) 
 \nonumber \\
+ (U_\lambda^i - \p_\lambda \bar\omega^i, g A_\lambda \times \omega_i). \ \ \ \ \ \ \ 
\eeqa
and a partially reduced set of variables is defined by
\beqa
\label{primesources}
{M'_\lambda}^i & \equiv & M_\lambda^i + \p_\lambda \bar\phi^i  
\nonumber  \\
N'_{\lambda i} & \equiv & N_{\lambda i} - \p_\lambda \omega_i
\nonumber  \\
{U'_\lambda}^i & \equiv & U_\lambda^i - \p_\lambda \bar\omega^i
\nonumber  \\
V'_{\lambda i} & \equiv & V_{\lambda i} + \p_\lambda \phi_i
\nonumber  \\
{K'_\lambda}^a & \equiv & K_\lambda^a - \p_\lambda \bar c^a - g (\bar\omega^i \times V_{\lambda i})^a
\nonumber  \\
&& \ \ \ \  - [ g (U_\lambda^i - \p_\lambda \bar\omega^i) \times \phi_i ]^a.
\eeqa
This gives the complete dependence of $\Gamma$ on $b, \bar c$ and on the 4 auxiliary ghosts $\phi, \omega, \bar\omega, \bar\phi$.

	There remains one ghost field $c$ which we have not yet considered and whose local equation of motion we cannot solve.  However it obeys an integrated equation of motion, which was first given for the present action in \cite{Schaden:1994}, that will be useful.  The operator,
\beqa
{\cal G}^a \equiv \int d^Dx \ \Big[ {\delta \over \delta c^a} + g f^{abd} \Big( i  \bar c^b {\delta \over \delta b^d} - \phi_i^b {\delta \over \delta \omega_i^d}
\nonumber \\
 - \bar\omega^{bi} {\delta \over \delta \bar\phi^{di}}
 + U_\lambda^{bi} {\delta \over \delta M_\lambda^{di}} + V_{\lambda i}^b {\delta \over \delta N_{\lambda i}^d} \Big) \Big],
\eeqa
is a symmetry of the local action $\Sigma$, apart from a breaking term that is linear in the local fields,
\beq
{\cal G}^a \Sigma = - g (K_\lambda \times A_\lambda)^a + g (L \times c)^a.
\eeq
The operator itself is also linear in the local fields, so the same identity holds for the quantum effective action
\beq
{\cal G}^a \Gamma = - g (K_\lambda \times A_\lambda)^a + g (L \times c)^a.
\eeq
We introduce
\beq
\label{spsv}
\Sigma_{inv} \equiv \Sigma_p + (K'_\lambda, g A_\lambda \times c) + (L, (-g/2)(c \times c)),
\eeq
which satisfies
\beq
{\cal G}^a \Sigma_{inv} = {\cal G}^a \Sigma.
\eeq
We separate this term out of the action and write,
\beq
\Gamma = \Sigma_{inv} + \Gamma_9(A, c, K', L, M', N', U', V'),
\eeq
so $\Gamma_9$ satisfies,
\beq
{\cal G}^a \Gamma_9 = {{\cal G}'}^a\Gamma_9 = 0,
\eeq
where
\beq
{ {\cal G}' }^a \equiv \int d^Dx \ \Big[ {\delta \over \delta c^a} + g f^{abd} \Big( {U'}_\lambda^{bi} {\delta \over \delta {M'}_\lambda^{di}} + {V'}_{\lambda i}^b {\delta \over \delta {N'}_{\lambda i}^d} \Big) \Big].
\eeq

	The new element in this Appendix, not found in \cite{Zwanziger:1993}, is the final change of variable
\beqa
\label{primetostar}
k_\lambda & \equiv & K'_\lambda
\nonumber \\
m_\lambda^{ai} & \equiv & {M'}_\lambda^{ai} - g ( c  \times {U'_\lambda}^i)^a
\nonumber \\
n_{\lambda i}^a & \equiv & {N'}_{\lambda i}^a - g ( c  \times V'_{\lambda i})^a
\nonumber \\
u_\lambda^{ai} & \equiv & {U_\lambda'}^{ai}
\nonumber \\
v_\lambda^{ai} & \equiv & {V'_\lambda}^{ai}
\eeqa
and the new fully reduced quantum effective action $\Gamma^*$ which is a functional of the new variables,
\beqa
\label{reducedstar}
\Gamma'(A, c, K', L, M', N', U', V') = \ \ \ \ \ \ \ \ \ \  \ \ \ \ \ \ \  
\nonumber \\
 (K'_\lambda, gA_\lambda \times c) 
+ (L, (-g/2) c \times c) \ \ \ \ 
\nonumber \\
+ \Gamma^*(A, c, k, L, m, n, u, v).
\eeqa
It satisfies the simple integrated ghost equation of motion
\beq
{\cal G}^{*a} \Gamma^* \equiv 
\int d^Dx \ {\delta \Gamma^* \over \delta c^a} = 0.
\eeq
This equation is equivalent to the statement that $\Gamma^*$ depends only on $\p_\mu c$, but not on $c$ itself, and we write
\beq
\label{partialc}
\Gamma^* = \Gamma^*(A, \p c, k, L, m, n, u, v).
\eeq
All other ghosts besides $c$ also appear in $\Gamma^*$ only as derivatives that are contained in the sources $k, m, n, u$ and $v$.  The fact that only derivatives of all ghost fields appears in $\Gamma^*$ is the functional expression of the well-known factorization of external ghost momenta from all vertices.

\section{One-loop calculation of the vacuum free energy}

	We wish to evaluate the dependence of the free energy $\Gamma(0, \gamma)$ upon $\gamma$, to one-loop order, starting from the action (1).  To this order, it is sufficient to consider the quadratic part of this action.  The quadratic action of the fermi-fields $c, \bar c, \omega, \bar\omega$ is independent of $\gamma$ and we suppose that they are integrated out.  There remains
\beqa
S_q & = & \int d^Dx \ [ (1/4)(\p_\mu A_\nu^a - \p_\nu A_\mu^a)^2 + i \p_\mu b^c A_\mu^c 
\nonumber \\
&&+ \p_\lambda \bar\phi_\mu^{ab} \p_\lambda \phi_\mu^{ab} + \gamma^{1/2} g f^{abc}A_\mu^b(\phi - \bar\phi)_\mu^{ca}].
\eeqa
More generally, since we are only interested in the $\gamma$-dependence, and $\gamma$ appears only in the $A$-$\phi$ and $A$-$\bar\phi$ mixing term, we may freely integrate out all other fields besides the ones that mix.  We integrate out the $b$ field.  This imposes the Landau-gauge constraint $\p \cdot A = 0$, so $A$ is purely transverse, $\p \cdot A = 0$.  We may decompose $\phi$ and $\bar\phi$ into their longitudinal and transverse parts.  Since $A$ is purely transverse, only the transverse parts of $\phi$ and $\bar\phi$ mix with~$A$.  We suppose that the longitudinal parts of $\phi$ and $\bar\phi$ are integrated out, so these fields are now also purely transverse.  We decompose $\phi$ and $\bar\phi$ according to
\beq
\phi = (1/\sqrt 2)(X + i Y); \ \ \ \ \   \bar\phi = (1/\sqrt 2)(X - i Y).
\eeq
Only $Y$ mixes with $A$, and we integrate out $X$.  The action now reads,
\beqa
S_q & = & \int d^Dx \ [ (1/2)(\p_\lambda A_\mu^a)^2  + (1/2) \p_\lambda Y_\mu^{ab} \p_\lambda Y_\mu^{ab}
\nonumber \\
&& \ \ \ \ \ \ \ \ \ \ \      + i \gamma^{1/2} g \sqrt 2 f^{abc}A_\mu^b Y_\mu^{ca}],
\eeqa
where is it understood that $A_\mu^b$ and $Y_\mu^b$ are both purely transverse.  Only the part of $Y_\mu^b$ that is projected onto the adjoint representation by
\beq
Y_\mu^a \equiv (1/\sqrt N) f^{abc} Y_\mu^{ca}
\eeq
mixes with $A$ and we integrate out the remaining components of $Y_\mu^b$, so the action simplifies to
\beqa
S_q & = & \int d^Dx \ [ (1/2) (\p_\lambda A_\mu^a)^2  + (1/2) (\p_\lambda Y_\mu^a)^2
\nonumber \\
&& \ \ \ \ \ \ \ \ \ \ \      + i \gamma^{1/2} g (2N)^{1/2} A_\mu^a Y_\mu^a].
\eeqa   
This is the correct normalization because
\beq
P^{ab, cd} \equiv (1/ N) \sum_e  f^{abe} f^{cde}
\eeq
is a projector
\beq
P^{ab,cd} P^{cd, ef} = P^{ab,ef}.
\eeq
We complete the diagonalization by forming the $i$-particles \cite{Baulieu:2009ha},
\beq
\lambda_\mu^b = (1/\sqrt 2) (A + Y)_\mu^b; \ \ \ \ \ \  \eta_\mu^b = (1/\sqrt 2) (A - Y)_\mu^b,
\eeq
so $S_q$ is diagonal,
\beqa
S_q & = & \int d^Dx \  (1/2)  [(\p_\mu \lambda_\nu^a)^2 + i M^2 (\lambda_\nu^a)^2
\nonumber \\
&& \ \ \ \ \ \ \ \    +  (\p_\mu \eta_\nu^a)^2 - i M^2 (\eta_\nu^a)^2],
\eeqa
where we have written
\beq
M^2 \equiv \gamma^{1/2} g (2N)^{1/2}.
\eeq

	This action is diagonal in momentum space.  We quantize in a periodic Euclidean box of volume $\Omega$, so
\beq
\lambda_\mu^a(x) = \Omega^{-1} \sum_k \exp(ik \cdot x) \ \lambda_\mu^a(k),
\eeq	
and similarly for $\eta$, so the action reads
\beqa
S_q & = &   (2\Omega)^{-1} \sum_k [ (k^2 + i M^2) |\lambda_\nu^a(k)|^2 
\nonumber \\
&& + ( k^2 - i M^2 ) |\eta_\nu^a(k)|^2  ],
\eeqa
where $k_\mu = 2\pi n_\mu/L$ and $\Omega = L^D$, and $n_\mu$ runs over all integers.  The one-loop contribution to the partition function, with all sources set to~0, is given by
\beq
Z_1(\gamma) = \int \prod d\lambda d\eta  \exp(- S_q)
\eeq
For each $k$ there are $N^2 - 1$ color components and $D -1$ (transverse) Lorentz components, of the fields $\lambda$ and $\eta$, which gives
\beq
Z_1(\gamma) = \prod_k [(k^2 + i M^2)(k^2 - i M^2)]^{- (N^2 -1)(D-1)/2}.
\eeq
With $Z_1(\gamma) = \exp[- \Gamma_1(\gamma)]$, we get
\beq
\Gamma_1(\gamma) = (N^2 -1)(D-1) {\rm Re} \sum_k \ln(k^2 + i M^2),
\eeq
or
\beq
{\Gamma_1(\gamma) \over (N^2 -1)(D-1) \Omega} =  {\rm Re} \int { d^Dk \over (2\pi)^D} \ln(k^2 + i M^2),
\eeq
where ${\rm Re}$ means real part.  We go to $D$-dimensional spherical coordinates to obtain
\beqa
{\Gamma_1(\gamma) \over \Omega} & = & (N^2 -1)(D-1) S_{D-1}
\\     \nonumber
&& \times {\rm Re} \int_0^\infty { dk^2 \ (k^2)^{(D-2)/2} \over 2 (2\pi)^D} \ln(k^2 + i M^2),
\eeqa
where $S_{D-1} = 2 \pi^{D/2}/\Gamma(D/2)$ is the area of a $D-1$ dimensional sphere.  With $y = k^2$ and dimensional regularization, this gives, after a partial integration,
\beq
{\Gamma_1(\gamma) \over \Omega} = -  {(N^2 -1)(D-1) S_{D-1} \over (2\pi)^D D } J
\eeq
where
\beq
J \equiv {\rm Re} \int_0^\infty dy \ { y^{D/2} \over y + i M^2 },
\eeq
or, with $y = iM^2 z$,
\beq
J \equiv {\rm Re} [\exp(i \pi /2) M^2]^{D/2} \int_0^\infty dz \ { z^{D/2} \over z + 1 }.
\eeq
This gives
\beqa
J & = & \cos(\pi D/4) M^D \int_0^\infty dz \int_0^\infty d\alpha \  z^{D/2} \exp[- (z+1)\alpha]
\nonumber \\
&= &  \cos(\pi D/4) M^D \Gamma(1 + D/2) \int_0^\infty d\alpha \ \alpha^{-(1 + D/2)}  \exp( - \alpha)
\nonumber \\
& = &  \cos(\pi D/4) M^D \Gamma(1 + D/2) \Gamma(- D/2)
\nonumber \\
& = &  \cos(\pi D/4) M^D { \pi \over \sin(- \pi D/2) }
\nonumber \\
& = & -    { \pi M^D \over 2 \sin(\pi D/4) }.
\eeqa
We obtain in dimension $D$,
\beq
{\Gamma_1(\gamma) \over \Omega} =  { (N^2 -1)(D-1) \pi M^D \over (4\pi)^{D/2} D  \Gamma(D/2) \sin[\pi(4 - D)/4] },
\eeq
or
\beq
{\Gamma_1(\gamma) \over \Omega} =  { (N^2 -1)(D-1) \pi  \over (4\pi)^{D/2} D  \Gamma(D/2) \sin[\pi(4 - D)/4] } (2N g^2 \gamma)^{D/4}.
\eeq
We write $D = 4 - \epsilon$, and take the limit $\epsilon \to 0$, which gives,
\beq
{\Gamma_1(\gamma) \over \Omega} =  {  3(N^2 -1) (2N g^2 \gamma) \over (4\pi)^2  } \Big( { 1 \over \epsilon } - {1 \over 4} \ln(2N g^2 \gamma/\mu^4) \Big),
\eeq
where we have introduced a normalization mass $\mu$.  We drop the pole term, and obtain for the one-loop contribution to the quantum effective action,
\beq
{\Gamma_1(\gamma) \over \Omega} = - {  3(N^2 -1) (2N g^2 \gamma)   \over 4 (4\pi)^2  } \ln(2N g^2 \gamma/\mu^4) .
\eeq

\section{Proof of identity}

In this Appendix we prove equality (\ref{2termsequal}).  By (\ref{partialQ}) it is sufficient to show
\beqa
\label{2termsequala}
 \int d^Dx \ \sum_{\lambda, a} { \delta W \over \delta {M}_{\lambda \lambda}^{aa} }\Big|_{\Phi = 0, Q = Q_1} \ \ \ \ \ \ \ \ \ \ \ \ \ \ \ \ \ 
\nonumber  \\
= - \int d^Dx \ \sum_{\lambda, a} { \delta W \over \delta {V}_{\lambda \lambda}^{aa} }\Big|_{\Phi = 0, Q = Q_1}.
\eeqa
In the formula for the partition function, we set to zero all sources besides $M$ and $V$, namely $K = L = N = U = 0$, and $J_\alpha = 0$ for all sources~$J_\alpha$ of the elementary fields~$\Phi_\alpha$.  We integrate out the $b$ field so the Landau gauge condition, $\p \cdot A = 0$, is satisfied on-shell.  We next integrate out the $\bar c$ field which gives $\delta({\cal M} c) = \det{\cal M} \ \delta(c)$, where ${\cal M} = - \p_\mu D_\mu = - D_\mu \p_\mu$ is the Faddeev-Popov operator which is hermitian because $\p \cdot A = 0$.  Here $\det{\cal M}$ is the Faddeev-Popov determinent, and $\delta(c)$ is the functional delta-function, which may be written in a mode expansion $\delta(c) = \prod_i c_i$.)  We next integrate out $c$, which results in setting $c = 0$ everywhere, and we integrate out the auxiliary fermi ghosts $\omega$ and $\bar\omega$ which gives a factor of $(\det {\cal M})^f$ .  As a result, the extended action $\Sigma$ is replaced by
\beqa
\Sigma_1 & = & (1/4) (F_{\mu \nu}, F_{\mu \nu}) + (\bar\phi_\mu^{ac}, {\cal M}^{ab} \phi_\mu^{bc})
\\   \nonumber
&& + (M_{\lambda \mu}^{ab}, D_\lambda \phi_\mu^{ab}) + (D_\lambda \bar\phi_\mu^{ab}, V_{\lambda \mu}^{ab}) 
 + (M_{\lambda \mu}^{ab}, V_{\lambda \mu}^{ab}),
\eeqa
and $Z = Z(M, V)$.  With $W(M, V) = \ln Z(M, V)$, we have
\beqa
{\delta W \over \delta M_{\lambda \mu}^{ab} }\Big|_{Q_1} & = & - Z^{-1} \int dA d\phi d\bar\phi \ (D_\lambda \phi_\mu^{ab} + V_{\lambda \mu}^{ab}) 
\nonumber \\
&& \ \ \ \ \ \ \ \ \ \ \ \ \ \ \ \ \ \ \ \ \     \times \exp( - \Sigma_1)|_{Q_1}
\nonumber \\ 
& = &- Z^{-1} \int dA d\phi d\bar\phi \ (D_\lambda \phi_\mu^{ab} - \gamma^{1/2} \delta_{\lambda \mu} \delta^{ab})
\nonumber \\
&& \ \ \ \ \ \ \ \ \ \ \ \ \ \ \ \ \ \ \ \ \ \ \  \times \exp( - \Sigma_2), 
\eeqa
where
\beqa
\Sigma_2 & \equiv & \Sigma_1|_{Q = Q_1}
\nonumber \\
&& = (1/4) (F_{\mu \nu}, F_{\mu \nu}) 
+ (\bar\phi_\mu^{ac}, {\cal M}^{ab} \phi_\mu^{bc})
\\   \nonumber
&&
+  \int d^Dx \ \{ \gamma^{1/2} [ D_\lambda (\phi - \bar\phi)_\lambda]^{aa} 
 - f \gamma \}.
\eeqa
Here we have set $M_{\lambda, \mu}^{ab} = - V_{\lambda, \mu}^{ab} = \gamma^{1/2} \delta_{\lambda, \mu} \delta^{ab}$ because $Q = Q_1$.  By a similar calculation we obtain
\beqa
{\delta W \over \delta V_{\lambda \mu}^{ab} }\Big|_{Q_1} 
& = &- Z^{-1} \int dA d\phi d\bar\phi \ (D_\lambda \bar\phi_\mu^{ab} + \gamma^{1/2} \delta_{\lambda \mu} \delta^{ab})
\nonumber \\
&& \ \ \ \ \ \ \ \ \ \ \ \ \ \ \ \ \ \ \ \ \ \ \  \times \exp( - \Sigma_2). 
\eeqa
In the last integral we make the change of variable $\phi = \bar\phi'$ and $\bar\phi = \phi'$.  Then, after dropping primes and using the hermiticity of the Faddeev-Popov operator, $(\phi_\mu^{ac}, {\cal M}^{ab} \bar\phi_\mu^{bc}) = (\bar\phi_\mu^{ac}, {\cal M}^{ab} \phi_\mu^{bc})$ which holds because $\p \cdot A = 0$, we obtain
\beqa
{\delta W \over \delta V_{\lambda \mu}^{ab} }\Big|_{Q_1} 
& = &- Z^{-1} \int dA d\phi d\bar\phi \ ( - D_\lambda \phi_\mu^{ab} + \gamma^{1/2} \delta_{\lambda \mu} \delta^{ab})
\nonumber \\
&& \ \ \ \ \ \ \ \ \ \ \ \ \ \ \ \ \ \ \ \ \ \ \  \times \exp( - \Sigma_2). 
\eeqa
This gives ${\delta W \over \delta V_{\lambda \mu}^{ab} }\Big|_{Q_1} = - {\delta W \over \delta M_{\lambda \mu}^{ab} }\Big|_{Q_1}$, which proves the assertion.

\appendix




\begin{thebibliography}{99}
\parskip=0pt



\bibitem{Gribov:1977wm}
V. N. Gribov,
Nucl. Phys. B {\bf 139}, 1978.


\bibitem{GOZ:2005}
J. Greensite, S. Olejnik, and D. Zwanziger, 
JHEP, 0505:070 (2005) and arXiv: hep-lat/0407032.


\bibitem{Zwanziger:1989mf}
  Daniel Zwanziger,
  Nucl.\ Phys.\  B {\bf 323}, 513 (1989).
  
  
\bibitem{Zwanziger:1993}
  Daniel Zwanziger,
  Nucl.\ Phys.\  B {\bf 399}, 477 (1993). 

\bibitem{Schaden:1994}
N. Maggiore and M. Schaden,
Phys. Rev. {\bf D50}, 6616 (1994).


\bibitem{Sorella:2005}
D. Dudal, R. F. Sobreiro, S. P. Sorella, H. Verschelde,
Phys. Rev. {\bf D72}, 014016 (2005) and arXiv: hep-th/0502183.


\bibitem{Gracey:0510}
J.A. Gracey,
Phys.~Lett.~ {\bf B632} 282 (2006) and hep-ph/0510151.
 
 
\bibitem{Gracey:0605} 
J.A. Gracey, 
JHEP 0605:052 (2006) and hep-ph/0605077.


\bibitem{Dudal:0806}
David Dudal, John A. Gracey, Silvio Paolo Sorella, Nele Vandersickel, Henri Verschelde, 
Phys.~Rev.~{\bf D78}, 065047 (2008) and arXiv:0806.4348 [hep-th].


\bibitem{Dudal:0808} 
D. Dudal, J.A. Gracey, S.P. Sorella, N. Vandersickel, H. Verschelde, 
Phys.~Rev.~{\bf D78}, 125012 (2008) and arXiv:0808.0893 [hep-th].


\bibitem{Sorella:0904}
D. Dudal, S.P. Sorella, N. Vandersickel, H. Verschelde,
Phys.~Rev.~{\bf D79}, 121701 (2009) and arXiv: 0904.0641 [hep-th].


\bibitem{Dudal:0908}
David Dudal, Silvio Paolo Sorella, Nele Vandersickel, Henri Verschelde, 
JHEP 0908:110 (2009) and arXiv:0906.4257 [hep-th].


\bibitem{Gracey:0909}
J.A. Gracey, 
JHEP, 1002:009 (2010) and arXiv:0909.3411 [hep-th].
 

\bibitem{Vandersickel0910}
N. Vandersickel, D. Dudal, S.P. Sorella, H. Verschelde, 
arXiv:0910.2653 [hep-th]


\bibitem{Dudal:0911}
D. Dudal, N. Vandersickel, H. Verschelde, S.P. Sorella, 
arXiv:0911.0082 [hep-th]

\bibitem{Baulieu:2009}
L. Baulieu, D. Dudal, M.S. Guimaraes, M.Q. Huber, S.P. Sorella, N. Vandersickel, D. Zwanziger, 
arXiv:0912.5153 [hep-th].


\bibitem{Dudal:1001}
D. Dudal, S.P. Sorella, N. Vandersickel, 
arXiv:1001.3103 [hep-th].


\bibitem{Nakajima:1978} 
T. Maskawa and H. Nakajima Prog. Theor. Phys. {\bf 60}, 1526 (1978), Prog. Theor. Phys. {\bf 63}, 642 (1980).


\bibitem{Semenov:1982}
M. Semenov-Tyan-Shanskii and V.  Franke,
Zap. Nauch. Sem. Leningrad. Otdelleniya Matematicheskogo Instituta in V. A. Steklov, AN SSSR, {\bf 120}, 159  (1982), (In English translation: New York, Plenum Press 1986).


\bibitem{Zwanziger:1982}
Daniel Zwanziger, 
Nucl. Phys. B {\bf 209} 336 (1982).

\bibitem{Sorella:0905}
S.P. Sorella
Phys.~Rev.~D {\bf 80} 025013 (2009) and arXiv:0905.1010.
  

\bibitem{Kugo-Ojima}
T. Kugo and I. Ojima, 
Prog. Theor. Phys. Suppl. {\bf 66} 1 (1979) [Erratum Prog. Theor. Phys. {\bf 71} 1121 (1984)].


\bibitem{Alkofer:2000}
Reinhard Alkofer and Lorenz von Smekal,
Physics Reports {\bf 353} 281 (2001) and arXiv:hep-ph/0007355.


\bibitem{Zwanziger:0904}
Daniel Zwanziger,
arXiv:0904.2380 [hep-th].


\bibitem{Huber:0910}
Markus Q. Huber, Reinhard Alkofer, Silvio P. Sorella,
arXiv:0910.5604 [hep-th].


\bibitem{Cucchieri:2010.03}
Attilio Cucchieri, Tereza Mendes
arXiv:1001.2584 [hep-lat].


\bibitem{Maas:2007} 
A.~Maas,
Phys. Rev. {\bf D75} 116004 (2007) and arXiv:0704.0722 [hep-lat]. 


\bibitem{Cucchieri:2008.12}
Attilio Cucchieri, Tereza Mendes,
 arXiv:0812.3261 [hep-lat].   

\bibitem{Cucchieri:2008.04}
Attilio Cucchieri, Tereza Mendes,
Phys.~Rev.~{\bf D78} (2008) 094503 and arXiv: 0804.2371 [hep-lat].


\bibitem{Ilgenfritz:2009}
I.L. Bogolubsky, E.-M. Ilgenfritz, M. MŸller-Preussker, A. Sternbeck,
arXiv:0901.0736 [hep-lat].


\bibitem{Oliveira:2008}
O. Oliveira, P. J. Silva,
arXiv:0809.0258 [hep-lat].
  


\bibitem{vonSmekal2007}
A. Sternbeck, L. von Smekal, D. B. Leinweber, A. G. Williams
PoS {\bf LAT2007} 340, 2007 arXiv:0710.1982 [hep-lat].
  
  
\bibitem{Cucchieri:2007b}  
A. Cucchieri, T. Mendes,
Phys. Rev Lett. {\bf 100} 241601, 2008 and arXiv:0712.3517 [hep-lat].


\bibitem{Maas:2008} Axel Maas,
Phys. Rev. {\bf D79}, 014505 (2009) and arXiv: 0808.3047 [hep-lat].



\bibitem[Dudal et~al.(2008)]{Dudal:2008sp}
D.~Dudal, J.~A. Gracey, S.~P. Sorella, N.~Vandersickel and H.~Verschelde,  \emph{Phys. Rev.} \textbf{D78}, 065047 (2008).


\bibitem[Baulieu et~al.(2009)]{Baulieu:2009ha}
L.~Baulieu, D.~Dudal, M.~S. Guimaraes, M.~Q. Huber, S.~P. Sorella,  N.~Vandersickel and D.~Zwanziger,  arXiv:0912.5153 [hep-th]  (2009).








\end{thebibliography}
\end{document}